\documentclass[preprint,showpacs,preprintnumbers,amsmath,amssymb]{revtex4}

\usepackage{graphicx}
\usepackage{dcolumn}
\usepackage{bm}
\usepackage{hyperref}
\hypersetup{colorlinks=false}

\begin{document}


\title{Four-Body Bound State Calculations in \\Three-Dimensional Approach}

\author{M.~R. Hadizadeh}
\email{hadizade@khayam.ut.ac.ir}
\author{S. Bayegan}%
 \email{bayegan@khayam.ut.ac.ir}
\affiliation{%
Department of Physics, University of Tehran, P.O.Box 14395-547,
Tehran, Iran
}%

\date{\today}

\begin{abstract}
The four-body bound state with two-body interactions is formulated
in Three-Dimensional approach, a recently developed momentum space
representation which greatly simplifies the numerical calculations
of few-body systems without performing the partial wave
decomposition. The obtained three-dimensional Faddeev-Yakubovsky
integral equations are solved with two-body spin-independent and
spin-averaged potentials. This is the first step toward the
calculations of four-nucleon bound state problem in
Three-Dimensional approach. Results for four-body binding energies
are in good agreement with achievements of the other methods.
\end{abstract}

\pacs{21.45.+v, 27.10.+h, 21.10.Dr, 21.30.-x }
\keywords{Suggested keywords}
\maketitle

\section{Introduction}
The bound state of few-body systems seems to be an ideal
laboratory to determine two-, three- and four-body nuclear forces.
The studies of the four-body bound state properties for the case
of few-body interactions have received increasing attention
theoretically and experimentally in recent years. Although the
four-body bound state poses a challenging problem numerically,
because of presence of fourth body, its investigation promises
insights into the rich structure of nuclear interactions. To this
aim one requires an accurate and reliable method to obtain the
full solution of four-body bound state in a straightforward
manner.

The four-body bound state calculations are carried out by
different methods to solve the nonrelativistic Schr\"{o}dinger
equation such as, the Coupled-Rearrangement-Channel Gaussian-basis
Variational(CRCGV) \cite{1}-\cite{7}, the Stochastic Variation(SV)
with correlated Gaussians \cite{8}-\cite{12}, the Hyperspherical
Harmonic variational (HH) \cite{13}-\cite{18}, the Green's
Function Monte Carlo(GFMC) \cite{19}-\cite{22}, the No-Core Shell
Model(NCSM) \cite{23}-\cite{28}, Effective Interaction
Hyperspherical Harmonic(EIHH) \cite{29},\cite{30} and the
Faddeev-Yakubovsky (F-Y). In the last method the nonrelativistic
schr\"{o}dinger equation is transformed  to two coupled sets of
finite number of coupled equations in three variables for the F-Y
amplitudes. The calculations based on F-Y are performed in
configuration space \cite{31}-\cite{34} and in momentum space
\cite{35}-\cite{43} after a partial wave (PW) expansion, where the
algebraic and algorithmic steps can be quiet involved. Though a
few partial waves often provide qualitative insight, modern
four-body calculations need 1572 or more different spin, isospin
and angular momentum combinations \cite{42},\cite{43}. It appears
therefore natural to avoid a PW representation completely and work
directly with vector variables. On this basis in recent years W.
Gl\"{o}ckle and collaborators have introduced the
Three-Dimensional (3D) approach which greatly simplifies the two-
and three-body scattering and bound state calculations without
using PW decomposition \cite{44}-\cite{53}.

Our aim in this paper is to extend this approach for four-body
bound state with two-body interactions, we work directly with
vector variables in the F-Y scheme in momentum space. Here the
dependence on momentum vectors, i.e. the magnitudes of momenta and
the angles between the momentum vectors, shows that the full
solution can be reached exactly and simply whereas the PW
representation of the amplitudes leads to rather complicated
expressions \cite{54}. The calculations of four-body bound state
with the three-body interactions is currently underway and it will
be reported elsewhere. As a simplification we neglect spin and
isospin degrees of freedom here and study the four-boson bound
state. So this work is the first step in the direction of solving
the four-nucleon bound state problem without performing the PW
decomposition.

Recently the four-boson bound state has been studied with
short-range forces and large scattering length at leading order in
an Effective Field Theory approach \cite{55}-\cite{57}, but this
investigation is also based on PW decomposition and the
interactions are restricted to only S-wave sector.

This paper is organized as follows. Section \ref{II} reviews the
F-Y equations for four-boson bound state. In section \ref{III} we
represent the coupled F-Y equations as function of momentum
vectors. In section \ref{IV} we discuss our choice for independent
variables of momentum and angle variables for the unknown
amplitudes in the equations and in their kernels, where this new
representation(3D) is contrasted with traditional PW
representation. Section \ref{V} describes details of our algorithm
for solving coupled F-Y three-dimensional integral equations. In
section \ref{VI} we compare our results for three- and four-boson
binding energies to results obtained from other techniques. In
order to test our calculation we investigate the stability of the
eigenvalue of the Yakubovsky kernel with respect to the number of
grid points and we calculate the expectation value of Hamiltonian
operator. Finally we summarize in section \ref{VII} and provide an
outlook.

\section{Four-Body Bound State Equations} \label{II}
The bound state of four identical particles which interact via
pairwise forces $V_{ij} (ij\equiv 12,\,13,\,14,\,23,\,24 $ and
$34)$ is given by Schr\"{o}dinger equation which reads in integral
form:
\begin{equation}
|\Psi\rangle=G_{0}\sum_{i<j} V_{ij}|\Psi\rangle \label{1}
\end{equation}
Here the free four-body propagator is given by
$G_{0}=(E-H_{0})^{-1}$, and $H_{0}$ stands for the free
hamiltonian. Introducing Yakubovsky components $|\Psi\rangle=
 \sum|\psi_{ij}\rangle $, with $|\psi_{ij}\rangle=G_{0}
V_{ij}|\Psi\rangle$ leads to the six coupled integral equations:
\begin{equation}
|\psi_{ij}\rangle=G_{0}t_{ij}\sum_{kl\neq ij} |\psi_{kl}\rangle
\label{2}
\end{equation}
The operator $t_{ij}$ describes the two-body $t-$matrix in the
two-body subsystem $ij$. We can rewrite Eq. (\ref{2}) as:
\begin{equation}
|\psi_{ij}\rangle=G_{0}t_{ij}(|\psi_{ik}\rangle +|\psi_{il}\rangle
+ |\psi_{jk}\rangle + |\psi_{jl}\rangle + |\psi_{kl}\rangle
)\label{3}
\end{equation}
Among various possibilities to decompose $|\psi_{ij}\rangle $ into
three F-Y components we choose the following one:
\begin{eqnarray}
|\psi_{ijk,l;ij}\rangle &=& G_{0}t_{ij}(|\psi_{ik}\rangle +
|\psi_{jk}\rangle ) \nonumber \\* |\psi_{ijl,k;ij}\rangle &=&
G_{0}t_{ij}(|\psi_{il}\rangle + |\psi_{jl}\rangle ) \nonumber
\\* |\psi_{ij,kl;ij}\rangle &=& G_{0}t_{ij}|\psi_{kl}\rangle
 \label{4}
\end{eqnarray}
The F-Y component
$|\psi_{ijk,l;ij}\rangle\,(|\psi_{ij,kl;ij}\rangle) $ belongs to a
$3+1\,(2+2)$ partition. They fulfill the following relation:
\begin{equation}
|\psi_{ij}\rangle=|\psi_{ijk,l;ij}\rangle +
|\psi_{ijl,k;ij}\rangle + |\psi_{ij,kl;ij}\rangle  \label{5}
\end{equation}
The multiple indices for the F-Y components denote the two-body
followed by the $3+1$ or $2+2$ fragmentation. It is easily seen
that every $|\psi_{ij}\rangle$ component contains two $3+1$ type
chains and one $2+2$ type chain, therefore total wave function
$|\Psi\rangle$ contains twelve different $3+1$ type chains and six
$2+2$ type chains. So altogether one has eighteen F-Y components.
If we consider identical particles (here bosons, since we are
omitting spin), the four-body wave function $|\Psi\rangle$ has to
be totally symmetric. As a consequence all twelve components of
$3+1$ type are identical in their functional form and only the
particles are permuted. The same is true for the six components of
$2+2$ type. Thus it is sufficient to consider only two independent
F-Y components corresponding to the $3+1$ and $2+2$ partitions,
\begin{eqnarray}
|\psi_{1}\rangle=|\psi_{123,4;12}\rangle \nonumber \\*
|\psi_{2}\rangle=|\psi_{12,34;12}\rangle \label{6}
\end{eqnarray}
After the straightforward derivation the $18$ coupled F-Y
components shrink to two coupled F-Y equations:
\begin{eqnarray}
|\psi_{1}\rangle=G_{0}t_{12}P[(1+P_{34})|\psi_{1}\rangle+|\psi_{2}\rangle]
\nonumber \\*
|\psi_{2}\rangle=G_{0}t_{12}\tilde{P}[(1+P_{34})|\psi_{1}\rangle+|\psi_{2}\rangle]
\label{7}
\end{eqnarray}
Where $P_{ij}$ is the permutation operator between the $i-th$ and
$j-th$ particle, and
\begin{eqnarray}
P &=& P_{12}P_{23}+P_{13}P_{23} \nonumber   \\* \tilde{P} &=&
P_{13}P_{24} \label{8}
\end{eqnarray}
The total four-body wave function is then given as:
\begin{equation}
|\Psi\rangle=(1+P+P_{34}P+\tilde{P})[(1+P_{34})|\psi_{1}\rangle+|\psi_{2}\rangle]
\label{9}
\end{equation}
The symmetry property of $|\psi_{1}\rangle$ under exchange of
particles $1$ and $2$, and $|\psi_{2}\rangle$ under separate
exchanges of particles $1,2$ and $3,4$ guarantee that
$|\Psi\rangle$ is totally symmetric.

We would like to add the remark that another derivation of F-Y
components is also possible \cite{35}. In this representation two
transition operators which follow the subcluster Faddeev-like
equations have been introduced as a function of two-body
transition operator $t_{12}$. Consequently the kernel of coupled
Yakubovsky integral equations contains two subcluster kernels that
should be evaluated by Pad\'{e} technique. So its numerical
calculations is more complicated and time consuming in comparison
to above derivation.

\section{Momentum Space Representation of Faddeev-Yakubovsky Equations}\label{III}
In order to solve the coupled equations (7), in momentum space we
introduce standard Jacobi momenta sets corresponding to both
$3+1\,(123,4;12)$ and $2+2\,(12,34;12)$ chains:
\begin{eqnarray}
\vec{u}_{1} &=& \frac{\vec{k}_{1}-\vec{k}_{2}}{2} \nonumber \\*
\vec{u}_{2} &=&
\frac{2}{3}(\vec{k}_{3}-\frac{\vec{k}_{1}+\vec{k}_{2}}{2})
\nonumber \\* \vec{u}_{3} &=&
\frac{3}{4}(\vec{k}_{4}-\frac{\vec{k}_{1}+\vec{k}_{2}+\vec{k}_{3}
}{3}) \nonumber \\*  \nonumber\\* \vec{v}_{1} &=&
\frac{\vec{k}_{1}-\vec{k}_{2}}{2} \nonumber \\* \vec{v}_{2} &=&
\frac{\vec{k}_{1}+\vec{k}_{2}}{2}-\frac{\vec{k}_{3}+\vec{k}_{4}}{2}
\nonumber \\* \vec{v}_{3}&=& \frac{\vec{k}_{3}-\vec{k}_{4}}{2}
 \label{10}
\end{eqnarray}
Then we introduce the four-body basis states corresponding to each
Jacobi momenta set:
\begin{eqnarray}
 &&|\vec{u}_{1}\,\vec{u}_{2}\,\vec{u}_{3}\rangle    \nonumber \\
 &&|\vec{v}_{1}\,\vec{v}_{2}\,\vec{v}_{3}\rangle \label{11}
\end{eqnarray}
Both basis states are complete in the four-body Hilbert space:
\begin{eqnarray}
\int D^{3}A \,\,|\vec{A}_{1}\,\vec{A}_{2}\,\vec{A}_{3}\rangle
\langle \vec{A}_{1}\,\vec{A}_{2}\,\vec{A}_{3}|=1  \label{12}
\end{eqnarray}
Where $\vec{A_{i}}$ indicates each one of $\vec{u}_{i}$ and
$\vec{v}_{i}$ vectors and $D^{3}A \equiv
d^{3}A_{1}\,d^{3}A_{2}\,d^{3}A_{3}$. Also they are normalized
according to:
\begin{equation}
\langle
\vec{A}_{1}\,\vec{A}_{2}\,\vec{A}_{3}|\vec{A}'_{1}\,\vec{A}'_{2}\,\vec{A}'_{3}\rangle=\delta^{3}(\vec{A}_{1}-\vec{A}'_{1})\,\delta^{3}(\vec{A}_{2}-\vec{A}'_{2})\,\delta^{3}(\vec{A}_{3}-\vec{A}'_{3})
\label{13}
\end{equation}
Clearly the basis states
$|\vec{u}_{1}\,\vec{u}_{2}\,\vec{u}_{3}\rangle $ are adequate to
expand F-Y component $|\psi_{1}\rangle$ and correspondingly the
basis states $|\vec{v}_{1}\,\vec{v}_{2}\,\vec{v}_{3}\rangle$ are
adequate for $|\psi_{2}\rangle$. Let us now represent coupled
equations, Eq. (\ref{7}), with respect to the basis states have
been introduced in  Eq. (\ref{11}):
\begin{eqnarray}
 \langle \vec{u}_{1}\,\vec{u}_{2}\,\vec{u}_{3}|\psi_{1}\rangle &=& \int
D^{3}u ''\,\langle\vec{u}_{1}\,\vec{u}_{2}\,\vec{u}_{3}|G_{0}t
P(1+P_{34})|\vec{u}\,''_{1}\,\vec{u}\,''_{2}\,\vec{u}\,''_{3}\rangle\langle\vec{u}\,''_{1}\,\vec{u}\,''_{2}\,\vec{u}\,''_{3}|\psi_{1}\rangle
 \nonumber  \nonumber \\*  &+& \int
D^{3}v'\,\langle\vec{u}_{1}\,\vec{u}_{2}\,\vec{u}_{3}|G_{0}t
P|\vec{v}\,'_{1}\,\vec{v}\,'_{2}\,\vec{v}\,'_{3}\rangle\langle\vec{v}\,'_{1}\,\vec{v}\,'_{2}\,\vec{v}\,'_{3}|\psi_{2}\rangle
 \nonumber \\* \nonumber \\*
\langle \vec{v}_{1}\,\vec{v}_{2}\,\vec{v}_{3}|\psi_{2}\rangle &=&
\int D^{3}u'\,\langle\vec{v}_{1}\,\vec{v}_{2}\,\vec{v}_{3}|G_{0}t
\tilde{P}(1+P_{34})|\vec{u}\,'_{1}\,\vec{u}\,'_{2}\,\vec{u}\,'_{3}\rangle\langle\vec{u}\,'_{1}\,\vec{u}\,'_{2}\,\vec{u}\,'_{3}|\psi_{1}\rangle
  \nonumber \\*  &+& \int
D^{3}v'\,\langle\vec{v}_{1}\,\vec{v}_{2}\,\vec{v}_{3}|G_{0}t
\tilde{P}|\vec{v}\,'_{1}\,\vec{v}\,'_{2}\,\vec{v}\,'_{3}\rangle\langle\vec{v}\,'_{1}\,\vec{v}\,'_{2}\,\vec{v}\,'_{3}|\psi_{2}\rangle
 \label{14}
\end{eqnarray}
It is convenient to insert again the completeness relations
between permutation operators, it results:
\begin{eqnarray}
 \langle \vec{u}_{1}\,\vec{u}_{2}\,\vec{u}_{3}|\psi_{1}\rangle &=& \int D^{3}u' \int
D^{3}u''\,\langle\vec{u}_{1}\,\vec{u}_{2}\,\vec{u}_{3}|G_{0}t P
|\vec{u}\,'_{1}\,\vec{u}\,'_{2}\,\vec{u}\,'_{3}\rangle \,
\nonumber
\\* && \hspace{5mm}
\times \langle\vec{u}\,'_{1}\,\vec{u}\,'_{2}\,\vec{u}\,'_{3}|
(1+P_{34})|\vec{u}\,''_{1}\,\vec{u}\,''_{2}\,
\vec{u}\,''_{3}\rangle\,\langle\vec{u}\,''_{1}\,\vec{u}\,''_{2}\,\vec{u}\,''_{3}|\psi_{1}\rangle
\nonumber \\*  &+& \int D^{3}u' \int D^{3}v'\,\
\langle\vec{u}_{1}\,\vec{u}_{2}\,\vec{u}_{3}|G_{0}t P
|\vec{u}\,'_{1}\,\vec{u}\,'_{2}\,\vec{u}\,'_{3}\rangle \nonumber
\\* && \hspace{5mm} \times
\langle\vec{u}\,'_{1}\,\vec{u}\,'_{2}\,\vec{u}\,'_{3}|\vec{v}\,'_{1}\,\vec{v}\,'_{2}\,\vec{v}\,'_{3}\rangle
\,\langle\vec{v}\,'_{1}\,\vec{v}\,'_{2}\,\vec{v}\,'_{3}|\psi_{2}\rangle
 \nonumber \\* \nonumber \\*
\langle \vec{v}_{1}\,\vec{v}_{2}\,\vec{v}_{3}|\psi_{2}\rangle &=&
\int D^{3}v' \int D^{3}u'\,
\langle\vec{v}_{1}\,\vec{v}_{2}\,\vec{v}_{3}|G_{0}t \tilde{P}|
\vec{v}\,'_{1}\,\vec{v}\,'_{2}\,\vec{v}\,'_{3}\rangle\, \nonumber
\\* && \hspace{5mm}
\times \langle\vec{v}\,'_{1}\,\vec{v}\,'_{2}\,\vec{v}\,'_{3}|
(1+P_{34})|\vec{u}\,'_{1}\,\vec{u}\,'_{2}\,\vec{u}\,'_{3}\rangle\langle\vec{u}\,'_{1}\,\vec{u}\,'_{2}\,\vec{u}\,'_{3}|\psi_{1}\rangle
  \nonumber \\*  &+& \int
D^{3}v'\,\langle\vec{v}_{1}\,\vec{v}_{2}\,\vec{v}_{3}|G_{0}t
\tilde{P}| \vec{v}\,'_{1}\,\vec{v}\,'_{2}\,\vec{v}\,'_{3}\rangle
\langle\vec{v}\,'_{1}\,\vec{v}\,'_{2}\,\vec{v}\,'_{3}|\psi_{2}\rangle
 \label{15}
\end{eqnarray}
For evaluating the coupled equations, Eq. (\ref{15}), we need to
evaluate the following matrix elements:
\begin{eqnarray}
&& \langle\vec{u}_{1}\,\vec{u}_{2}\,\vec{u}_{3}|G_{0}t P
|\vec{u}\,'_{1}\,\vec{u}\,'_{2}\,\vec{u}\,'_{3}\rangle
 \label{16} \\*
&& \langle\vec{v}_{1}\,\vec{v}_{2}\,\vec{v}_{3}|G_{0}t \tilde{P}|
\vec{v}\,'_{1}\,\vec{v}\,'_{2}\,\vec{v}\,'_{3}\rangle \label{17}
\\* && \langle\vec{u}\,'_{1}\,\vec{u}\,'_{2}\,\vec{u}\,'_{3}|
(1+P_{34})|\vec{u}\,''_{1}\,\vec{u}\,''_{2}\,
\vec{u}\,''_{3}\rangle \label{18} \\* &&
\langle\vec{v}\,'_{1}\,\vec{v}\,'_{2}\,\vec{v}\,'_{3}|
(1+P_{34})|\vec{u}\,'_{1}\,\vec{u}\,'_{2}\,\vec{u}\,'_{3}\rangle
\label{19}
\end{eqnarray}
For evaluating the first term, Eq. (\ref{16}), we should insert
again a completeness relation between the two-body $t-$matrix
operator and permutation operator $P$ as:
\begin{eqnarray}
&& \langle\vec{u}_{1}\,\vec{u}_{2}\,\vec{u}_{3}|G_{0}t P
|\vec{u}\,'_{1}\,\vec{u}\,'_{2}\,\vec{u}\,'_{3}\rangle =
\frac{1}{E-\frac{u_{1}^{2}}{m}
-\frac{3u_{2}^{2}}{4m}-\frac{2u_{3}^{2}}{3m}} \nonumber
\\* && \hspace{10mm} \times \int D^{3}u''
\langle\vec{u}_{1}\,\vec{u}_{2}\,\vec{u}_{3}|t
|\vec{u}\,''_{1}\,\vec{u}\,''_{2}\,\vec{u}\,''_{3}\rangle
 \langle\vec{u}\,''_{1}\,\vec{u}\,''_{2}\,\vec{u}\,''_{3}|
P|\vec{u}\,'_{1}\,\vec{u}\,'_{2}\, \vec{u}\,'_{3}\rangle
\label{20}
\end{eqnarray}
Where the matrix elements of two-body $t-$matrix and permutation
operator $P$ are evaluated separately as:
\begin{eqnarray}
\langle\vec{u_{1}}\,\vec{u_{2}}\,\vec{u_{3}}|t
|\vec{u}\,''_{1}\,\vec{u}\,''_{2}\,\vec{u}\,''_{3}\rangle &=&
\delta^{3}(\vec{u_{2}}-\vec{u}\,''_{2})\,
\delta^{3}(\vec{u_{3}}-\vec{u}\,''_{3})\,
\langle\vec{u_{1}}|t(\epsilon) |\vec{u}\,''_{1}\rangle \nonumber
\\*  \epsilon &=& E-\frac{3u_{2}^{2}}{4m}-\frac{2u_{3}^{2}}{3m} \label{21}
\end{eqnarray}
\begin{eqnarray}
\langle\vec{u}\,''_{1}\,\vec{u}\,''_{2}\,\vec{u}\,''_{3}|
P|\vec{u}\,'_{1}\,\vec{u}\,'_{2}\, \vec{u}\,'_{3}\rangle &=&
\delta^{3}(\vec{u}\,''_{3}-\vec{u}\,'_{3}) \nonumber
\\* &\times& \{\,\, \delta^{3}(\vec{u}\,''_{1}+\frac{1}{2}\vec{u}\,'_{1}-
\frac{3}{4}\vec{u}\,'_{2}) \, \delta^{3}(\vec{u}\,''_{2}
+\vec{u}\,'_{1}+ \frac{1}{2}\vec{u}\,'_{2} )
 \, \nonumber \\* && \,\, +\, \delta^{3}(\vec{u}\,''_{1}+\frac{1}{2}\vec{u}\,'_{1}+
\frac{3}{4}\vec{u}\,'_{2}) \, \delta^{3}(\vec{u}\,''_{2}
-\vec{u}\,'_{1}+ \frac{1}{2}\vec{u}\,'_{2} ) \,\,\} \nonumber
\\* \label{22}
\end{eqnarray}
For evaluation the matrix elements of permutation operator $P$ we
have used the relation between Jacobi momenta in different
two-body subsystems $(312,4;12),(231,4;12)$ and $(123,4;12)$.
Inserting Eqs. (\ref{21}) and $(\ref{22})$ into Eq. (\ref{20})
leads to:
\begin{eqnarray}
\langle\vec{u}_{1}\,\vec{u}_{2}\,\vec{u}_{3}|G_{0}t P
|\vec{u}\,'_{1}\,\vec{u}\,'_{2}\,\vec{u}\,'_{3}\rangle &=&
\frac{\delta^{3}(\vec{u}_{3}-\vec{u}\,'_{3})}{E-\frac{u_{1}^{2}}{m}
-\frac{3u_{2}^{2}}{4m}-\frac{2u_{3}^{2}}{3m}} \nonumber
\\* &&   \{ \,\, \delta^{3}(\vec{u}_{2}
+\vec{u}\,'_{1}+ \frac{1}{2}\vec{u}\,'_{2} )\,
\langle\vec{u}_{1}|t(\epsilon) |\frac{1}{2}\vec{u}_{2}
+\vec{u}\,'_{2}\rangle \nonumber
\\* && \,\,+ \delta^{3}(\vec{u}_{2}
-\vec{u}\,'_{1}+ \frac{1}{2}\vec{u}\,'_{2} )\,
\langle\vec{u}_{1}|t(\epsilon) |\frac{-1}{2}\vec{u}_{2}
-\vec{u}\,'_{2}\rangle  \} \nonumber
\\* \label{23}
\end{eqnarray}
Representation of the second term, Eq. (\ref{17}), follows the
similar steps:
\begin{eqnarray}
\langle\vec{v}_{1}\,\vec{v}_{2}\,\vec{v}_{3}|G_{0}t \tilde{P}
|\vec{v}\,'_{1}\,\vec{v}\,'_{2}\,\vec{v}\,'_{3}\rangle &=&
\frac{1}{E-\frac{v_{1}^{2}}{m}
-\frac{v_{2}^{2}}{2m}-\frac{v_{3}^{2}}{m}} \int D^{3}v''
\langle\vec{v}_{1}\,\vec{v}_{2}\,\vec{v}_{3}|t
|\vec{v}\,''_{1}\,\vec{v}\,''_{2}\,\vec{v}\,''_{3}\rangle
\nonumber
\\* &\times&
\langle\vec{v}\,''_{1}\,\vec{v}\,''_{2}\,\vec{v}\,''_{3}|
\tilde{P}|\vec{v}\,'_{1}\,\vec{v}\,'_{2}\, \vec{v}\,'_{3}\rangle
 \label{24}
\end{eqnarray}
The matrix elements of two-body $t-$matrix and permutation
operator $\tilde{P}$ are evaluated as:
\begin{eqnarray}
\langle\vec{v}_{1}\,\vec{v}_{2}\,\vec{v}_{3}|t
|\vec{v}\,''_{1}\,\vec{v}\,''_{2}\,\vec{v}\,''_{3}\rangle &=&
\delta^{3}(\vec{v}_{2}-\vec{v}\,''_{2})\,
\delta^{3}(\vec{v}_{3}-\vec{v}\,''_{3})\,
\langle\vec{v}_{1}|t(\epsilon^{*}) |\vec{v}\,''_{1}\rangle
\nonumber
\\*  \epsilon^{*}&=&E-\frac{v_{2}^{2}}{2m}-\frac{v_{3}^{2}}{m}
\label{25}
\end{eqnarray}
\begin{eqnarray}
\langle\vec{v}\,''_{1}\,\vec{v}\,''_{2}\,\vec{v}\,''_{3}|
\tilde{P}|\vec{v}\,'_{1}\,\vec{v}\,'_{2}\, \vec{v}\,'_{3}\rangle
&=& \delta^{3}(\vec{v}\,''_{1}-\vec{v}\,'_{3}) \,
\delta^{3}(\vec{v}\,''_{2}+\vec{v}\,'_{2}) \,
\delta^{3}(\vec{v}\,''_{3} -\vec{v}\,'_{1}) \label{26}
\end{eqnarray}
Inserting Eqs. (\ref{25}) and (\ref{26}) into Eq. (\ref{24}) leads
to:
\begin{eqnarray}
\langle\vec{v}_{1}\,\vec{v}_{2}\,\vec{v}_{3}|G_{0}t \tilde{P}
|\vec{v}\,'_{1}\,\vec{v}\,'_{2}\,\vec{v}\,'_{3}\rangle &=&
\frac{\delta^{3}(\vec{v}_{2}+\vec{v}\,'_{2})\,
\delta^{3}(\vec{v}_{3}-\vec{v}\,'_{1}) }{E-\frac{v_{1}^{2}}{m}
-\frac{v_{2}^{2}}{2m}-\frac{v_{3}^{2}}{m}} \,\,
\langle\vec{v}_{1}|t(\epsilon^{*}) |\vec{v}\,'_{3}\rangle
\label{27}
\end{eqnarray}
For evaluation the third term, Eq. (\ref{18}), we should use the
relation between Jacobi momenta in different chains $(123,4 ;12)$
and $(124,3;12)$, which leads to:
\begin{eqnarray}
\langle\vec{u}\,'_{1}\,\vec{u}\,'_{2}\,\vec{u}\,'_{3}|
(1+P_{34})|\vec{u}\,''_{1}\,\vec{u}\,''_{2}\,
\vec{u}\,''_{3}\rangle &=&
\delta^{3}(\vec{u}\,'_{1}-\vec{u}\,''_{1})\nonumber \\* &\times&
\{\,\,
\delta^{3}(\vec{u}\,'_{2}-\vec{u}\,''_{2})\,\delta^{3}(\vec{u}\,'_{3}-\vec{u}\,''_{3})
\nonumber \\* && \,\, \,+  \delta^{3}(\vec{u}\,'_{2}
-\frac{1}{3}\vec{u}\,''_{2} -\frac{8}{9}\vec{u}\,''_{3} )\,
\delta^{3}(\vec{u}\,'_{3} -\vec{u}\,''_{2}
+\frac{1}{3}\vec{u}\,''_{3} )\,\} \nonumber
\\* \label{28}
\end{eqnarray}
And finally for evaluation the fourth term, Eq. (\ref{19}), we
should use the relation between Jacobi momenta in two naturally
different chains $(123,4;12)$ and $(12,34;12)$, which leads to:
\begin{eqnarray}
\langle\vec{v}\,'_{1}\,\vec{v}\,'_{2}\,\vec{v}\,'_{3}|
(1+P_{34})|\vec{u}\,'_{1}\,\vec{u}\,'_{2}\, \vec{u}\,'_{3}\rangle
&=& \delta^{3}(\vec{u}\,'_{1}-\vec{v}\,'_{1})\nonumber \\*
&\times& \{\,\,
\delta^{3}(\vec{u}\,'_{2}+\frac{2}{3}\vec{v}\,'_{2}-
\frac{2}{3}\vec{v}\,'_{3} )\,\
\delta^{3}(\vec{u}\,'_{3}+\frac{1}{2}\vec{v}\,'_{2}+\vec{v}\,'_{3})
\nonumber
\\* && \,\, \,+  \delta^{3}(\vec{u}\,'_{2}+\frac{2}{3}\vec{v}\,'_{2}+
\frac{2}{3}\vec{v}\,'_{3} )\,\
\delta^{3}(\vec{u}\,'_{3}+\frac{1}{2}\vec{v}\,'_{2}-\vec{v}\,'_{3})\,\}
\nonumber
\\*  \label{29}
\end{eqnarray}
Finally inserting Eqs. (\ref{23}), (\ref{27}), (\ref{28}) and
(\ref{29}) in Eq. (\ref{15}) yields:
\begin{eqnarray}
\langle \vec{u}_{1}\,\vec{u}_{2}\,\vec{u}_{3}|\psi_{1}\rangle &=&
\frac{1}{{E-\frac{u_{1}^{2}}{m}
-\frac{3u_{2}^{2}}{4m}-\frac{2u_{3}^{2}}{3m}}}\int d^{3}u_{2}'
\,\, \langle\vec{u}_{1}|t_{s}(\epsilon) |\frac{1}{2}\vec{u}_{2}
+\vec{u}\,'_{2}\rangle \, \nonumber \\* &\times& \{\,\,
\langle\vec{u}_{2}+\frac{1}{2} \vec{u}\,'_{2} \,\,
\vec{u}\,'_{2}\,\,\vec{u}_{3}|\psi_{1}\rangle \nonumber \nonumber
\\* \quad && \hspace{2mm} +\langle\vec{u}_{2}+\frac{1}{2}
\vec{u}\,'_{2} \,\, \frac{1}{3}\vec{u}\,'_{2}+
\frac{8}{9}\vec{u}_{3} \,\,
\vec{u}\,'_{2}-\frac{1}{3}\vec{u}_{3}|\psi_{1}\rangle
 \nonumber \nonumber \\*  && \hspace{2mm} +
 \langle\vec{u}_{2}+\frac{1}{2} \vec{u}\,'_{2}\,\, -\vec{u}\,'_{2}-\frac{2}{3}\vec{u}_{3} \,\,
\frac{1}{2}\vec{u}\,'_{2}-\frac{2}{3}\vec{u}_{3}
|\psi_{2}\rangle\,\, \}
 \nonumber \\* \nonumber \\*
\langle \vec{v}_{1}\,\vec{v}_{2}\,\vec{v}_{3}|\psi_{2}\rangle &=&
\frac{\frac{1}{2}}{E-\frac{v_{1}^{2}}{m}
-\frac{v_{2}^{2}}{2m}-\frac{v_{3}^{2}}{m}} \,\,  \int d^{3}v_{3}'
\, \langle\vec{v}_{1}|t_{s}(\epsilon^{*})| \vec{v}\,'_{3}\rangle\,
\nonumber \\* &\times& \{\,\, 2\, \langle\vec{v}_{3}\,\,
 \frac{2}{3}\vec{v}_{2}+\frac{2}{3}\vec{v}\,'_{3} \,\, \frac{1}{2}\vec{v}_{2}-\vec{v}\,'_{3} |\psi_{1}\rangle
  \nonumber \\*  && \hspace{2mm}+
\langle\vec{v}_{3}\,\,-\vec{v}_{2}\,\,
\vec{v}\,'_{3}|\psi_{2}\rangle \,\,\} \label{30}
 \end{eqnarray}
Here $\langle\vec{a}|t_{s}(\varepsilon)| \vec{b}\rangle\,$
generally represents the symmetrized two-body $t-$matrix which is
defined as,
\begin{equation}
\langle\vec{a}|t_{s}(\varepsilon)| \vec{b}\rangle\,=\langle
\vec{a}|t(\varepsilon)| \vec{b}\rangle\, + \langle
\vec{a}|t(\varepsilon)| -\vec{b}\rangle\, \label{31}
\end{equation}
We would like to mention that the so obtained F-Y amplitudes
fulfill the below symmetry relations, as can be seen from Eq.
(\ref{30}):
\begin{eqnarray}
\langle \vec{u}_{1}\,\vec{u}_{2}\,\vec{u}_{3}|\psi_{1}\rangle &=&
\langle -\vec{u}_{1}\,\vec{u}_{2}\,\vec{u}_{3}|\psi_{1}\rangle
\nonumber \\*
\langle\vec{v}_{1}\,\vec{v}_{2}\,\vec{v}_{3}|\psi_{2}\rangle
\nonumber &=&
\langle-\vec{v}_{1}\,\vec{v}_{2}\,\vec{v}_{3}|\psi_{2}\rangle
\nonumber
\\* \langle\vec{v}_{1}\,\vec{v}_{2}\,\vec{v}_{3}|\psi_{2}\rangle
&=& \langle\vec{v}_{1}\,\vec{v}_{2}\,-\vec{v}_{3}|\psi_{2}\rangle
\label{32}
\end{eqnarray}
From the F-Y components $|\psi_{1} \rangle$ and $|\psi_{2}
\rangle$ the four-body wave function is obtained by adding the
components defined in different $3+1$ and $2+2$ type chains as
given in Eq. (\ref{9}). After evaluating the permutation operators
$P, \, \tilde{P}$ and $P_{34}$ the wave function is given as:
\begin{equation}
|\Psi \rangle=|\Psi_{1} \rangle+|\Psi_{2} \rangle \label{33}
\end{equation}
Where $|\Psi_{1} \rangle\,(|\Psi_{2} \rangle)$ is corresponding to
all $3+1\,(2+2)$ type chains:
\begin{eqnarray}
\langle \vec{u}_{1}\,\, \vec{u}_{2}\,\,
\vec{u}_{3}|\Psi_{1}\rangle &=& \nonumber \\*
 && \hspace{-30mm} \{\,\, \langle
\vec{u}_{1}\,\,\vec{u}_{2}\,\,\vec{u}_{3}|\psi_{1}\rangle\,
\nonumber
\\* && \hspace{-30mm} \quad+ \langle
\frac{-1}{2}\vec{u}_{1}-\frac{3}{4}\vec{u}_{2}\,\,\,\vec{u}_{1}-\frac{1}{2}\vec{u}_{2}\,\,\,\vec{u}_{3}|\psi_{1}\rangle\,
\nonumber \\* && \hspace{-30mm} \quad+ \langle
\frac{-1}{2}\vec{u}_{1}+\frac{3}{4}\vec{u}_{2}\,\,\,-\vec{u}_{1}-\frac{1}{2}\vec{u}_{2}\,\,\,\vec{u}_{3}|\psi_{1}\rangle\,
\}^{\, 123+4} \nonumber \\*
 &&\hspace{-30mm} \,\,+\{\,\, \langle
\vec{u}_{1}\,\, \frac{1}{3}\vec{u}_{2}+\frac{8}{9}\vec{u}_{3} \,\,
\vec{u}_{2}-\frac{1}{3}\vec{u}_{3}|\psi_{1}\rangle\, \nonumber
\\* && \hspace{-30mm} \quad+ \langle
\frac{-1}{2}\vec{u}_{1}-\frac{1}{4}\vec{u}_{2}-\frac{2}{3}\vec{u}_{3}
\,\,\,\vec{u}_{1}-\frac{1}{6}\vec{u}_{2}-\frac{4}{9}\vec{u}_{3}
\,\,\, \vec{u}_{2}-\frac{1}{3}\vec{u}_{3}|\psi_{1}\rangle\,
\nonumber \\* &&\hspace{-30mm} \quad+ \langle
\frac{-1}{2}\vec{u}_{1}+\frac{1}{4}\vec{u}_{2}+\frac{2}{3}\vec{u}_{3}
\,\,\,-\vec{u}_{1}-\frac{1}{6}\vec{u}_{2}-\frac{4}{9}\vec{u}_{3}
\,\,\, \vec{u}_{2}-\frac{1}{3}\vec{u}_{3}|\psi_{1}\rangle\, \}^{\,
124+3} \nonumber \\*
 &&\hspace{-30mm} \,\,+\{\,\, \langle
\frac{1}{2}\vec{u}_{1}-\frac{3}{4}\vec{u}_{2}\,\,
-\frac{1}{3}\vec{u}_{1}-\frac{1}{6}\vec{u}_{2}+\frac{8}{9}
\vec{u}_{3} \,\,
-\vec{u}_{1}-\frac{1}{2}\vec{u}_{2}-\frac{1}{3}\vec{u}_{3}|\psi_{1}\rangle\,
\nonumber
\\* &&\hspace{-30mm} \quad+ \langle
\frac{1}{2}\vec{u}_{2}-\frac{2}{3}\vec{u}_{3}
\,\,\,\frac{2}{3}\vec{u}_{1}-\frac{2}{3}\vec{u}_{2}-\frac{4}{9}\vec{u}_{3}
\,\,\,
-\vec{u}_{1}-\frac{1}{2}\vec{u}_{2}-\frac{1}{3}\vec{u}_{3}|\psi_{1}\rangle\,
\nonumber \\* &&\hspace{-30mm} \quad+ \langle
\frac{-1}{2}\vec{u}_{1}+\frac{1}{4}\vec{u}_{2}+\frac{2}{3}\vec{u}_{3}
\,\,\,-\frac{1}{3}\vec{u}_{1}+\frac{5}{6}\vec{u}_{2}-\frac{4}{9}\vec{u}_{3}
\,\,\,
-\vec{u}_{1}-\frac{1}{2}\vec{u}_{2}-\frac{1}{3}\vec{u}_{3}|\psi_{1}\rangle\,
\}^{\, 134+2} \nonumber \\*
 &&\hspace{-30mm} \,\,+\{\,\, \langle
-\frac{1}{2}\vec{u}_{1}-\frac{3}{4}\vec{u}_{2}\,\,
\frac{1}{3}\vec{u}_{1}-\frac{1}{6}\vec{u}_{2}+\frac{8}{9}
\vec{u}_{3} \,\,
\vec{u}_{1}-\frac{1}{2}\vec{u}_{2}-\frac{1}{3}\vec{u}_{3}|\psi_{1}\rangle\,
\nonumber
\\* &&\hspace{-30mm} \quad+ \langle
\frac{1}{2}\vec{u}_{2}-\frac{2}{3}\vec{u}_{3}
\,\,\,-\frac{2}{3}\vec{u}_{1}-\frac{2}{3}\vec{u}_{2}-\frac{4}{9}\vec{u}_{3}
\,\,\,
\vec{u}_{1}-\frac{1}{2}\vec{u}_{2}-\frac{1}{3}\vec{u}_{3}|\psi_{1}\rangle\,
\nonumber \\* &&\hspace{-30mm} \quad+ \langle
\frac{1}{2}\vec{u}_{1}+\frac{1}{4}\vec{u}_{2}+\frac{2}{3}\vec{u}_{3}
\,\,\,\frac{1}{3}\vec{u}_{1}+\frac{5}{6}\vec{u}_{2}-\frac{4}{9}\vec{u}_{3}
\,\,\,
\vec{u}_{1}-\frac{1}{2}\vec{u}_{2}-\frac{1}{3}\vec{u}_{3}|\psi_{1}\rangle\,
\}^{\, 234+1} \label{34}
\end{eqnarray}

\begin{eqnarray}
\langle \vec{v}_{1}\,\, \vec{v}_{2}\,\,
\vec{v}_{3}|\Psi_{2}\rangle &=& \nonumber \\*
 && \hspace{-30mm} \,\, \{\,\, \langle
\vec{v}_{1}\,\,\vec{v}_{2}\,\,\vec{v}_{3}|\psi_{2}\rangle\,
\nonumber
\\* && \hspace{-30mm} \quad+  \langle
\vec{v}_{3}\,\,\,-\vec{v}_{2}\,\,\,\vec{v}_{1}|\psi_{2}\rangle\,
\}^{\, 12+34} \nonumber \\*
 && \hspace{-30mm} \,\,+\{\,\, \langle
\frac{1}{2}\vec{v}_{1}+\frac{1}{2}\vec{v}_{2}-\frac{1}{2}\vec{v}_{3}
\,\,\,\vec{v}_{1}+\vec{v}_{3} \,\,\,
-\frac{1}{2}\vec{v}_{1}+\frac{1}{2}\vec{v}_{2}+\frac{1}{2}\vec{v}_{3}|\psi_{2}\rangle\,
\nonumber
\\* && \hspace{-30mm} \quad+ \langle
\frac{-1}{2}\vec{v}_{1}+\frac{1}{2}\vec{v}_{2}+\frac{1}{2}\vec{v}_{3}
\,\,\,-\vec{v}_{1}-\vec{v}_{3} \,\,\,
\frac{1}{2}\vec{v}_{1}+\frac{1}{2}\vec{v}_{2}-\frac{1}{2}\vec{v}_{3}|\psi_{2}\rangle\,
\}^{\, 13+24} \nonumber \\*
 && \hspace{-30mm} \,\,+\{\,\, \langle
\frac{1}{2}\vec{v}_{1}+\frac{1}{2}\vec{v}_{2}
+\frac{1}{2}\vec{v}_{3}\,\,\, \vec{v}_{1}-\vec{v}_{3} \,\,\,
-\frac{1}{2}\vec{v}_{1}+\frac{1}{2}\vec{v}_{2}-\frac{1}{2}\vec{v}_{3}|\psi_{2}\rangle\,
\nonumber
\\* && \hspace{-30mm} \quad+ \langle
-\frac{1}{2}\vec{v}_{1}+\frac{1}{2}\vec{v}_{2}-\frac{1}{2}\vec{v}_{3}
\,\,\,-\vec{v}_{1}+\vec{v}_{3} \,\,\,
\frac{1}{2}\vec{v}_{1}+\frac{1}{2}\vec{v}_{2}+\frac{1}{2}\vec{v}_{3}|\psi_{2}\rangle\,
\}^{\, 14+23} \label{35}
\end{eqnarray}
Each curly bracket contains all possible chains in the subsystem
which is indicated with corresponding superscript. Already here we
see that:
\begin{eqnarray}
\langle \vec{u}_{1}\,\vec{u}_{2}\,\vec{u}_{3}|\Psi_{1}\rangle &=&
\langle -\vec{u}_{1}\,\vec{u}_{2}\,\vec{u}_{3}|\Psi_{1}\rangle
\nonumber \\*
\langle\vec{v}_{1}\,\vec{v}_{2}\,\vec{v}_{3}|\Psi_{2}\rangle
\nonumber &=&
\langle-\vec{v}_{1}\,\vec{v}_{2}\,\vec{v}_{3}|\Psi_{2}\rangle
\nonumber
\\* \langle\vec{v}_{1}\,\vec{v}_{2}\,\vec{v}_{3}|\Psi_{2}\rangle
&=& \langle\vec{v}_{1}\,\vec{v}_{2}\,-\vec{v}_{3}|\Psi_{2}\rangle
\label{36}
\end{eqnarray}
Eq. (\ref{36}) is satisfied if the F-Y components fulfill the
expected symmetries in Eq. (\ref{32}).

\section{Choosing The Coordinate Systems}\label{IV}
\begin{figure}
\includegraphics{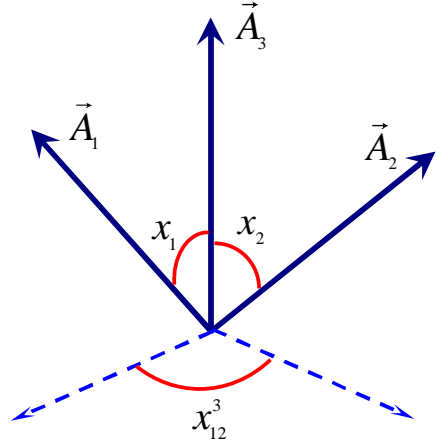}
\caption{\label{figure 1} The geometry of three vectors
$\vec{A}_{1},\, \vec{A}_{2}$ and $\vec{A}_{3}$ relevant in the
four-body bound state problem. The independent angle variables
$x_{1}$, $x_{2}$ and $x_{12}^{3}$ as defined in Eq. (\ref{37}) are
indicated. The dashed arrows represent the normal vectors
$(\vec{A}_{3} \times \vec{A}_{1})$ and $(\vec{A}_{3} \times
\vec{A}_{2})$.}
\end{figure}

The F-Y components
$|\psi_{i}(\vec{A}_{1}\,\,\vec{A}_{2}\,\,\vec{A}_{3} ) \rangle$
are given as a function of Jacobi momenta vectors and as a
solution of coupled three-dimensional integral equations, Eq.
(\ref{30}). Since we ignore spin and isospin dependencies, the
both F-Y components
$|\psi_{i}(\vec{A}_{1}\,\,\vec{A}_{2}\,\,\vec{A}_{3})\rangle$ are
scalars and thus only depend on the magnitudes of Jacobi momenta
and the angles between them. The first important step for an
explicit calculation is the selection of independent variables.
One needs six variables to uniquely specify the geometry of the
three vectors $\vec{A}_{1},\, \vec{A}_{2}$ and $\vec{A}_{3}$,
which are shown in Fig. \ref{figure 1}. Having in mind that with
three vectors one can span 2 planes, i.e. the
$\vec{A}_{3}-\vec{A}_{1}$ plane and $\vec{A}_{3}-\vec{A}_{2}$
plane, a natural choice of independent variables is \cite{53}:
\begin{eqnarray}
&& A_{1}=|\vec{A}_{1}| \quad A_{2}=|\vec{A}_{2}| \quad
A_{3}=|\vec{A}_{3}| \nonumber
\\*  && x_{1}=\hat{\vec{A}}_{3}.\hat{\vec{A}}_{1} \quad x_{2}=\hat{\vec{A}}_{3}.\hat{\vec{A}}_{2} \quad
x_{12}^{3}=( \widehat{\vec{A}_{3}\times \vec{A}_{1}}
).(\widehat{\vec{A}_{3}\times \vec{A}_{2}} ) \label{37}
\end{eqnarray}

The last variable, $x_{12}^{3}$, is the angle between the two
normal vectors of the $\vec{A}_{3}-\vec{A}_{1}$ plane and the
$\vec{A}_{3}-\vec{A}_{2}$ plane, which is explicitly related to
the angle between $\vec{A}_{1}$ and $\vec{A}_{2}$ vectors as:
\begin{equation}
x_{12}^{3}=\frac{\hat{\vec{A}}_{1}.\hat{\vec{A}}_{2}
-(\hat{\vec{A}}_{3}.\hat{\vec{A}}_{1})(\hat{\vec{A}}_{3}.\hat{\vec{A}}_{2})}{\sqrt{1-(\hat{\vec{A}}_{3}.\hat{\vec{A}}_{1})^{2}}\sqrt{1-(\hat{\vec{A}}_{3}.\hat{\vec{A}}_{2})^{2}}}
\label{38}
\end{equation}
Therefore in order to solve Eq. (\ref{30}) directly without
employing PW projection, we have to define suitable coordinate
systems. As shown in Fig. \ref{figure 2}, for both F-Y components
we choose the third vector parallel to $Z-$axis, the second vector
in the $X-Z$ plane and express the remaining vectors, the first as
well as the integration vectors, with respect to them. We have the
magnitudes of vectors as well as the following angle relations as
variables:

\begin{figure}
\includegraphics{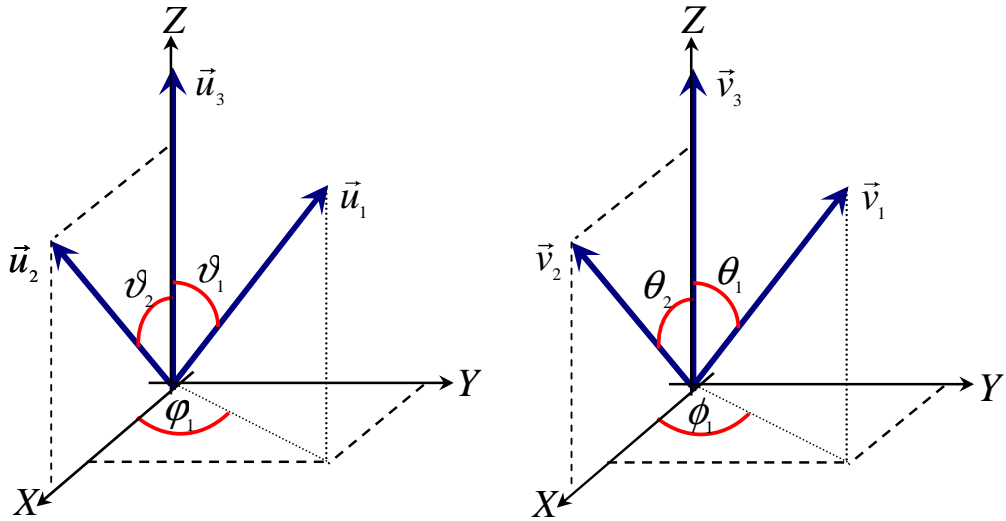}
\caption{\label{figure 2} These figures show the geometry of both
vector sets $(\vec{u}_{1},\, \vec{u}_{2},\,\vec{u}_{3})$ and
$(\vec{v}_{1},\, \vec{v}_{2},\,\vec{v}_{3})$ relevant in the F-Y
componets. The third vectors $\vec{u}_{3}, \, \vec{v}_{3}$ have
been chosen parallel to $Z-$axis, the second vectors $\vec{u}_{2},
\, \vec{v}_{2}$ in the $X-Z$ plane and the first vectors
$\vec{u}_{1}, \, \vec{v}_{1}$ are free in the space.
 The independent
angle variables $x_{1}$, $x_{2}, x_{12}^{3}$ and $X_{1}$, $X_{2},
X_{12}^{3}$ as defined in Eqs. (\ref{39}) and (\ref{40}) are
indicated.}
\end{figure}

\begin{eqnarray}
 x_{1}&=&\hat{\vec{u}}_{3}.\hat{\vec{u}}_{1}\equiv
\cos(\vartheta_{1})\nonumber\\*
x_{2}&=&\hat{\vec{u}}_{3}.\hat{\vec{u}}_{2} \equiv
\cos(\vartheta_{2})\nonumber\\* x_{12}^{3}&=&
\hat{\vec{u}}_{1}^{xy}.\hat{\vec{u}}_{2}^{xy} \equiv
\cos(\varphi_{1})\nonumber\\*  y_{12}&=&
\hat{\vec{u}}_{1}.\hat{\vec{u}}_{2} \equiv
x_{1}x_{2}+\sqrt{1-x_{1}^{2}}\sqrt{1-x_{2}^{2}}\cos(\varphi_{1})\nonumber\\*
x_{2}'&=&\hat{\vec{u}}_{3}.\hat{\vec{u}}'_{2} \equiv
\cos(\vartheta_{2}^{'})\nonumber\\*
y_{12^{'}}&=&\hat{\vec{u}}_{1}.\hat{\vec{u}}'_{2} \equiv
x_{1}x_{2}'+\sqrt{1-x_{1}^{2}}\sqrt{1-x_{2}'^{2}}\cos(\varphi_{1}-\varphi'_{2})\nonumber\\*
 y_{22^{'}}&=&\hat{\vec{u}}_{2}.\hat{\vec{u}}'_{2} \equiv
x_{2}x_{2}'+\sqrt{1-x_{2}^{2}}\sqrt{1-x_{2}'^{2}}\cos(\varphi'_{2})
\label{39}
\end{eqnarray}
\begin{eqnarray}
X_{1}&=&\hat{\vec{v}}_{3}.\hat{\vec{v}}_{1} \equiv
\cos(\theta_{1})\nonumber\\*
X_{2}&=&\hat{\vec{v}}_{3}.\hat{\vec{v}}_{2} \equiv
\cos(\theta_{2})\nonumber\\*
X_{12}^{3}&=&\hat{\vec{v}}_{1}^{xy}.\hat{\vec{v}}_{2}^{xy} \equiv
\cos(\phi_{1})\nonumber\\*
Y_{12}&=&\hat{\vec{v}}_{1}.\hat{\vec{v}}_{2} \equiv
X_{1}X_{2}+\sqrt{1-X_{1}^{2}}\sqrt{1-X_{2}^{2}}\cos(\phi_{1})\nonumber\\*
X_{3}'&=&\hat{\vec{v}}_{3}.\hat{\vec{v}}'_{3} \equiv
\cos(\theta_{3}^{'})\nonumber\\*
Y_{13^{'}}&=&\hat{\vec{v}}_{1}.\hat{\vec{v}}'_{3} \equiv
X_{1}X_{3}'+\sqrt{1-X_{1}^{2}}\sqrt{1-X_{3}'^{2}}\cos(\phi_{1}-
\phi'_{3})\nonumber\\*
Y_{23^{'}}&=&\hat{\vec{v}}_{2}.\hat{\vec{v}}'_{3} \equiv
X_{2}X_{3}'+\sqrt{1-X_{2}^{2}}\sqrt{1-X_{3}'^{2}}\cos(\phi'_{3})
\label{40}
\end{eqnarray}

With this choice of variables the matrix elements of F-Y
components are given as:
\begin{eqnarray}
 \langle \vec{u}_{1}\,\, \vec{u}_{2}\,\,
\vec{u}_{3}|\psi_{1}\rangle & \equiv &
\psi_{1}(u_{1}\,u_{2}\,u_{3}\,x_{1}\,x_{2}\, x_{12}^{3} )
\nonumber\\*  \langle \vec{v}_{1}\,\, \vec{v}_{2}\,\,
\vec{v}_{3}|\psi_{2}\rangle &\equiv&
\psi_{2}(v_{1}\,v_{2}\,v_{3}\,X_{1}\,X_{2}\, X_{12}^{3} )
\label{41}
\end{eqnarray}
Furthermore $\langle\vec{u}_{1}|t_{s}(\epsilon)
|\frac{1}{2}\vec{u}_{2} +\vec{u}\,'_{2}\rangle$ and
$\langle\vec{v}_{1}|t_{s}(\epsilon^{*})| \vec{v}\,'_{3}\rangle$
are also scalar functions, and then can be written in the
following form:
\begin{eqnarray}
 \langle\vec{u}_{1}|t_{s}(\epsilon) |\frac{1}{2}\vec{u}_{2}
+\vec{u}\,'_{2}\rangle &\equiv&
t_{s}(\vec{u}_{1},\frac{1}{2}\vec{u}_{2} +\vec{u}\,'_{2};\epsilon)
\equiv t_{s}(u_{1},\tilde{\pi},\tilde{x} ;\epsilon) \nonumber\\*
\langle\vec{v}_{1}|t_{s}(\epsilon^{*})| \vec{v}\,'_{3}\rangle
&\equiv& t_{s}(\vec{v}_{1},\vec{v}\,'_{3};\epsilon^{*}) \equiv
t_{s}(v_{1},v_{3}',Y_{13^{'}};\epsilon^{*}) \label{42}
\end{eqnarray}
Where
\begin{eqnarray}
 \tilde{\pi}&=&|\frac{1}{2}\vec{u}_{2}
+\vec{u}\,'_{2}|=\sqrt{\frac{1}{4}u_{2}^{2}+u_{2}'^{2}+u_{2}u_{2}'y_{22^{'}}}
\nonumber\\*
\tilde{x}&=&\hat{\vec{u}}_{1}.(\widehat{\frac{1}{2}\vec{u_{2}}+\vec{u}\,'_{2}}
)=\frac{1}{\tilde{\pi}}(\frac{1}{2}u_{2}y_{12}+u_{2}'y_{12^{'}})
\label{43}
\end{eqnarray}
The more complex dependencies appear under the integrals in Eq.
(\ref{30}) for magnitude and angle variables of F-Y components.
According to Eq. (\ref{37}) and Eq. (\ref{38}) they are given as:
\begin{eqnarray}
&& \langle \vec{u}_{2}+\frac{1}{2} \vec{u}\,'_{2} \,\,
\vec{u}\,'_{2}\,\,\vec{u}_{3}|\psi_{1}\rangle \equiv
\psi_{1}(\pi_{1}\,\, u_{2}'\,\, {u}_{3}\,\,
x_{12}\,\,x_{13}\,\,x_{\pi_{1}u_{2}'}^{u_{3}})
 \nonumber \\* &&
\langle\vec{u}_{2}+\frac{1}{2} \vec{u}\,'_{2} \,\,
\frac{1}{3}\vec{u}\,'_{2}+ \frac{8}{9}\vec{u}_{3} \,\,
\vec{u}\,'_{2}-\frac{1}{3}\vec{u}_{3}|\psi_{1}\rangle \equiv
\psi_{1}(\pi_{1}\,\, \pi_{2}\,\, \pi_{3}\,\,
x_{22}\,\,x_{23}\,\,x_{\pi_{1}\pi_{2}}^{\pi_{3}})
 \nonumber \nonumber \\*  &&
 \langle\vec{u}_{2}+\frac{1}{2} \vec{u}\,'_{2}\,\, -\vec{u}\,'_{2}-\frac{2}{3}\vec{u}_{3} \,\,
\frac{1}{2}\vec{u}\,'_{2}-\frac{2}{3}\vec{u}_{3} |\psi_{2}\rangle
\equiv \psi_{2}(\pi_{1}\,\, \pi_{4}\,\, \pi_{5}\,\,
x_{32}\,\,x_{33}\,\,x_{\pi_{1}\pi_{4}}^{\pi_{5}})
 \nonumber \\*
&&  \langle\vec{v}_{3}\,\,
 \frac{2}{3}\vec{v}_{2}+\frac{2}{3}\vec{v}\,'_{3} \,\, \frac{1}{2}\vec{v}_{2}-\vec{v}\,'_{3}
 |\psi_{1}\rangle \equiv \psi_{1}(v_{3}\,\,\Sigma_{1} \,\,\Sigma_{2} \,\,
X_{12}\,\,X_{13}\,\,X_{v_{3}\Sigma_{1}}^{\Sigma_{2}})
  \nonumber \\*  &&
\langle\vec{v}_{3}\,\,-\vec{v}_{2}\,
\vec{v}\,'_{3}|\psi_{2}\rangle \equiv \psi_{2}(v_{3}\,\,v_{2}
\,\,v_{3}' \,\, X_{22}\,\,X_{23}\,\,X_{v_{3}v_{2}}^{v_{3}'})
\label{44}
 \end{eqnarray}
Where the shifted arguments are:
\begin{eqnarray}
\pi_{1}&=&|\vec{u}_{2}+\frac{1}{2}
\vec{u}\,'_{2}|=\sqrt{u_{2}^{2}+\frac{1}{4}u_{2}'^{2}+u_{2}u_{2}'y_{22^{'}}}
\nonumber
\\* \pi_{2}&=&|\frac{1}{3}\vec{u}\,'_{2}+ \frac{8}{9}\vec{u}_{3}|=\sqrt{\frac{1}{9}u_{2}'^{2}+\frac{64}{81}u_{3}^{2}+\frac{16}{27}u_{2}'u_{3}x_{2}'} \nonumber \\*
\pi_{3}&=&|\vec{u}\,'_{2}-
\frac{1}{3}\vec{u}_{3}|=\sqrt{u_{2}'^{2}+\frac{1}{9}u_{3}^{2}-\frac{2}{3}u_{2}'u_{3}x_{2}'}
\nonumber \\*   \pi_{4}&=&|-\vec{u}\,'_{2}-
\frac{2}{3}\vec{u}_{3}|=\sqrt{u_{2}'^{2}+\frac{4}{9}u_{3}^{2}+\frac{4}{3}u_{2}'u_{3}x_{2}'}
\nonumber \\*  \pi_{5}&=&|\frac{1}{2}\vec{u}\,'_{2}-
\frac{2}{3}\vec{u}_{3}|=\sqrt{\frac{1}{4}u_{2}'^{2}+\frac{4}{9}u_{3}^{2}-\frac{2}{3}u_{2}'u_{3}x_{2}'}
\nonumber\\*
\Sigma_{1}&=&|\frac{2}{3}\vec{v}_{2}+\frac{2}{3}\vec{v}\,'_{3}|=\frac{2}{3}\sqrt{v_{2}^{2}+v_{3}'^{2}+2v_{2}v_{3}'Y_{23^{'}}}
\nonumber \\*
\Sigma_{2}&=&|\frac{1}{2}\vec{v}_{2}-\vec{v}\,'_{3}|=\sqrt{\frac{1}{4}v_{2}^{2}+v_{3}'^{2}-v_{2}v_{3}'Y_{23^{'}}}
\label{45}
\end{eqnarray}
\begin{eqnarray}
 x_{11}&=&(\widehat{\vec{u}_{2}+\frac{1}{2}\vec{u}\,'_{2}} )
.\hat{\vec{u}}_{2}'
=\frac{1}{\pi_{1}}(u_{2}y_{22^{'}}+\frac{1}{2}u_{2}') \nonumber
\\*   x_{12}&=&(\widehat{\vec{u}_{2}+\frac{1}{2}\vec{u}\,'_{2}}
) .\hat{\vec{u}}_{3}
=\frac{1}{\pi_{1}}(u_{2}x_{2}+\frac{1}{2}u_{2}'x_{2}')
 \nonumber
\\*  x_{13}&=&\hat{\vec{u}}_{2}'. \hat{\vec{u}}_{3}=x_{2}' \nonumber \\*
 x_{\pi_{1}u_{2}'}^{u_{3}} &=&\frac{x_{11}-x_{12}x_{13}}{\sqrt{1-x_{12}^{2}}\sqrt{1-x_{13}^{2}}}
\label{46}
\end{eqnarray}
\begin{eqnarray}
 x_{21}&=&(\widehat{\vec{u}_{2}+\frac{1}{2}\vec{u}\,'_{2}} )
.(\widehat{\frac{1}{3}\vec{u}\,'_{2}+\frac{8}{9}\vec{u}_{3}} )
\nonumber
\\*  &=& \frac{1}{\pi_{1}\pi_{2}}(\frac{1}{3}u_{2}
u_{2}'y_{22^{'}}+\frac{8}{9}u_{2}u_{3}x_{2}+\frac{1}{6}u_{2}'^{2}
+\frac{4}{9}u_{2}'u_{3}x_{2}') \nonumber
\\*  x_{22}&=&(\widehat{\vec{u}_{2}+\frac{1}{2}\vec{u}\,'_{2}} )
.(\widehat{\vec{u}\,'_{2}-\frac{1}{3}\vec{u}_{3}} ) \nonumber
\\* &=& \frac{1}{\pi_{1}\pi_{3}}(u_{2}
u_{2}'y_{22^{'}}-\frac{1}{3}u_{2}u_{3}x_{2}+\frac{1}{2}u_{2}'^{2}
-\frac{1}{6}u_{2}'u_{3}x_{2}')
 \nonumber
\\*  x_{23}&=&(\widehat{\frac{1}{3}\vec{u}\,'_{2}+\frac{8}{9}\vec{u}_{3}}
). (\widehat{\vec{u}\,'_{2}-\frac{1}{3}\vec{u}_{3}} ) \nonumber
\\*
&=& \frac{1}{\pi_{2}\pi_{3}}(\frac{1}{3}
u_{2}'+\frac{7}{9}u_{2}'u_{3}x_{2}'-\frac{8}{27}u_{3}^{2} )
\nonumber
\\*  x_{\pi_{1}\pi_{2}}^{\pi_{3}}
&=&
\frac{x_{21}-x_{22}x_{23}}{\sqrt{1-x_{22}^{2}}\sqrt{1-x_{23}^{2}}}
\label{47}
\end{eqnarray}
\begin{eqnarray}
x_{31}&=&(\widehat{\vec{u}_{2}+\frac{1}{2}\vec{u}\,'_{2}} )
.(\widehat{-\vec{u}\,'_{2}-\frac{2}{3}\vec{u}_{3}} ) \nonumber
\\*  &=&
\frac{-1}{\pi_{1}\pi_{4}}(u_{2}
u_{2}'y_{22^{'}}+\frac{2}{3}u_{2}u_{3}x_{2}+\frac{1}{2}u_{2}'^{2}
+\frac{1}{3}u_{2}'u_{3}x_{2}') \nonumber
\\*  x_{32}&=& (\widehat{\vec{u}_{2}+\frac{1}{2}\vec{u}\,'_{2}} )
.(\widehat{\frac{1}{2}\vec{u}\,'_{2}-\frac{2}{3}\vec{u}_{3}} )
\nonumber
\\* &=&
\frac{1}{\pi_{1}\pi_{5}}(\frac{1}{2}u_{2}
u_{2}'y_{22^{'}}-\frac{2}{3}u_{2}u_{3}x_{2}+\frac{1}{4}u_{2}'^{2}
-\frac{1}{3}u_{2}'u_{3}x_{2}')
 \nonumber
\\* x_{33}&=& (\widehat{-\vec{u}\,'_{2}-\frac{2}{3}\vec{u}_{3}}
). (\widehat{\frac{1}{2}\vec{u}\,'_{2}-\frac{2}{3}\vec{u}_{3}} )
\nonumber
\\*  &=& \frac{-1}{\pi_{4}\pi_{5}}(\frac{1}{2}
u_{2}'-\frac{1}{3}u_{2}'u_{3}x_{2}'-\frac{4}{9}u_{3}^{2} )
\nonumber
\\* x_{\pi_{1}\pi_{4}}^{\pi_{5}}
 &=&
\frac{x_{31}-x_{32}x_{33}}{\sqrt{1-x_{32}^{2}}\sqrt{1-x_{33}^{2}}}
\label{48}
\end{eqnarray}
\begin{eqnarray}
X_{11}&=&\hat{\vec{v}}_{3}.(\widehat{\frac{2}{3}\vec{v}_{2}+\frac{2}{3}\vec{v}\,'_{3}}
) =\frac{\frac{2}{3}}{\Sigma_{1}}(v_{2}X_{2}+v_{3}'X_{3}')
\nonumber
\\*  X_{12}&=&\hat{\vec{v}}_{3}.(\widehat{\frac{1}{2}\vec{v}_{2}-\vec{v}\,'_{3}}
) =\frac{1}{\Sigma_{2}}(\frac{1}{2}v_{2}X_{2}-v_{3}'X_{3}')
\nonumber
 \nonumber
\\* X_{13}&=&(\widehat{\frac{2}{3}\vec{v}_{2}+\frac{2}{3}\vec{v}\,'_{3}}
).(\widehat{\frac{1}{2}\vec{v}_{2}-\vec{v}\,'_{3}} )
=\frac{\frac{2}{3}}{\Sigma_{1}\Sigma_{2}}(\frac{1}{2}v_{2}^{2}-\frac{1}{2}v_{2}v_{3}'Y_{23^{'}}-v_{3}'^{2})
\nonumber
\\* X_{v_{3}\Sigma_{1}}^{\Sigma_{2}}
&=&\frac{X_{11}-X_{12}X_{13}}{\sqrt{1-X_{12}^{2}}\sqrt{1-X_{13}^{2}}}
\label{49}
\end{eqnarray}
\begin{eqnarray}
 X_{21}&=&\hat{\vec{v}}_{3}.(-\hat{\vec{v}}_{2}) =-X_{2} \nonumber
\\*   X_{22}&=&\hat{\vec{v}}_{3}.\hat{\vec{v}}'_{3}=X_{3}'
\nonumber
 \nonumber
\\* X_{23}&=&(-\hat{\vec{v}}_{2}).\hat{\vec{v}}'_{3}=-Y_{23^{'}}
\nonumber
\\* X_{v_{3}v_{2}}^{v_{3}'}
&=&\frac{X_{21}-X_{22}X_{23}}{\sqrt{1-X_{22}^{2}}\sqrt{1-X_{23}^{2}}}
\label{50}
\end{eqnarray}
These considerations lead to the explicit representation for the
F-Y components $|\psi_{1}\rangle $ and $|\psi_{2}\rangle$:

\begin{eqnarray}
\psi_{1}(u_{1}\,u_{2}\,u_{3}\,x_{1}\,x_{2}\, x_{12}^{3} ) &=&
\frac{1}{{E-\frac{u_{1}^{2}}{m}
-\frac{3u_{2}^{2}}{4m}-\frac{2u_{3}^{2}}{3m}}}\int_{0}^{\infty}
du'_{2} \, u'^{2}_{2} \int_{-1}^{+1} dx'_{2} \int_{0}^{2\pi}
d\varphi'_{2} \nonumber \\* && t_{s}(u_{1},\tilde{\pi},\tilde{x}
;\epsilon) \, \nonumber \\* &\times& \{\,\, \psi_{1}(\pi_{1}\,\,
u_{2}'\,\, {u}_{3}\,\,
x_{12}\,\,x_{13}\,\,x_{\pi_{1}u_{2}'}^{u_{3}}) \nonumber
\\* \quad && \hspace{2mm} +\psi_{1}(\pi_{1}\,\, \pi_{2}\,\, \pi_{3}\,\,
x_{22}\,\,x_{23}\,\,x_{\pi_{1}\pi_{2}}^{\pi_{3}})
 \nonumber \\*  && \hspace{2mm} +
 \psi_{2}(\pi_{1}\,\, \pi_{4}\,\, \pi_{5}\,\,
x_{32}\,\,x_{33}\,\,x_{\pi_{1}\pi_{4}}^{\pi_{5}})\,\,\}
 \nonumber \\* \nonumber \\*
\psi_{2}(v_{1}\,v_{2}\,v_{3}\,X_{1}\,X_{2}\, X_{12}^{3} ) &=&
\frac{\frac{1}{2}}{E-\frac{v_{1}^{2}}{m}
-\frac{v_{2}^{2}}{2m}-\frac{v_{3}^{2}}{m}} \,\,  \int_{0}^{\infty}
dv'_{3} \, v'^{2}_{3} \int_{-1}^{+1} dX_{3}' \int_{0}^{2\pi}
d\phi'_{3} \nonumber \\* &&
t_{s}(v_{1},v_{3}',Y_{13^{'}};\epsilon^{*})
 \nonumber \\* &\times& \{\,\, 2\,
\psi_{1}(v_{3}\,\,\Sigma_{1} \,\,\Sigma_{2} \,\,
X_{12}\,\,X_{13}\,\,X_{v_{3}\Sigma_{1}}^{\Sigma_{2}})
  \nonumber \\*  && \hspace{2mm} + \psi_{2}(v_{3}\,\,v_{2}
\,\,v_{3}' \,\, X_{22}\,\,X_{23}\,\,X_{v_{3}v_{2}}^{v_{3}'})
\,\,\} \label{51}
 \end{eqnarray}
The above coupled equations, Eq. (\ref{51}), is the starting point
for numerical calculations, and the details will be described in
the next section. In a standard PW representation Eq. (\ref{15})
is replaced by two coupled sets of finite number of coupled
integral equations \cite{40}:

\begin{eqnarray}
 \langle u_{1}\, u_{2}\, u_{3} \, \alpha_{1}|\psi_{1}\rangle &=& \sum_{\alpha'_{1}}  \int D^{3}u' \sum_{\alpha''_{1}} \int
D^{3}u'' \sum_{\alpha'''_{1}}  \int D^{3}u''' \nonumber \\* &&
\langle u_{1}\, u_{2}\, u_{3} \, \alpha_{1}|G_{0}t |u'_{1}\,
u'_{2}\, u'_{3} \, \alpha'_{1} \rangle \, \langle u'_{1}\,
u'_{2}\, u'_{3} \, \alpha'_{1}| P| u''_{1}\, u''_{2}\, u''_{3} \,
\alpha''_{1} \rangle \nonumber \\* && \langle u''_{1}\, u''_{2}\,
u''_{3} \, \alpha''_{1}|(1+P_{34})| u'''_{1}\, u'''_{2}\, u'''_{3}
\, \alpha'''_{1} \rangle  \langle u'''_{1}\, u'''_{2}\, u'''_{3}
\, \alpha'''_{1} |\psi_{1}\rangle
  \nonumber \\*  &+& \sum_{\alpha'_{1}}  \int D^{3}u' \sum_{\alpha''_{1}} \int
D^{3}u'' \sum_{\alpha'_{2}} \int D^{3}v' \nonumber \\* && \langle
u_{1}\, u_{2}\, u_{3} \, \alpha_{1} |G_{0}t |u'_{1}\, u'_{2}\,
u'_{3} \, \alpha'_{1} \rangle \,  \langle u'_{1}\, u'_{2}\, u'_{3}
\, \alpha'_{1}| P| u''_{1}\, u''_{2}\, u''_{3} \, \alpha''_{1}
\rangle \nonumber \\* && \langle u''_{1}\, u''_{2}\, u''_{3} \,
\alpha''_{1} | v'_{1} \, v'_{2} \, v'_{3} \, \alpha'_{2} \rangle
\,\langle v'_{1} \, v'_{2} \, v'_{3} \, \alpha'_{2}
|\psi_{2}\rangle
 \nonumber \\*  \nonumber \\*
\langle v_{1}\, v_{2}\, v_{3} \, \alpha_{2} |\psi_{2}\rangle &=&
\sum_{\alpha'_{2}} \int D^{3}v' \sum_{\alpha''_{2}} \int D^{3}v''
\sum_{\alpha'_{1}} \int D^{3}u'\, \nonumber \\* && \langle v_{1}\,
v_{2}\, v_{3} \, \alpha_{2}|G_{0}t| v'_{1} \, v'_{2} \, v'_{3} \,
\alpha'_{2} \rangle\, \langle v'_{1} \, v'_{2} \, v'_{3} \,
\alpha'_{2}| \tilde{P}| v''_{1} \, v''_{2} \, v''_{3} \,
\alpha''_{2} \rangle \nonumber \\* && \langle v''_{1} \, v''_{2}
\, v''_{3} \, \alpha''_{2}| u'_{1}\, u'_{2}\, u'_{3} \,
\alpha'_{1} \rangle  \langle u'_{1}\, u'_{2}\, u'_{3} \,
\alpha'_{1}|(1+P_{34})\psi_{1}\rangle
  \nonumber \\*  &+& \sum_{\alpha'_{2}} \int
D^{3}v' \sum_{\alpha''_{2}} \int D^{3}v'' \,\langle v_{1}\,
v_{2}\, v_{3} \, \alpha_{2}|G_{0}t | v'_{1} \, v'_{2} \, v'_{3} \,
\alpha'_{2}\rangle \nonumber \\* && \langle v'_{1} \, v'_{2} \,
v'_{3} \, \alpha'_{2}|\tilde{P}| v''_{1} \, v''_{2} \, v''_{3} \,
\alpha''_{2} \rangle \langle v''_{1} \, v''_{2} \, v''_{3} \,
\alpha''_{2}| \psi_{2}\rangle
 \label{52}
\end{eqnarray}
Where $\alpha_{1}\equiv(l_{1}l_{2})j_{3}, (j_{3}l_{3});J=0$ and
$\alpha_{2}\equiv(\lambda_{1}\lambda_{2})I, (I \lambda_{3});J=0$.
Here the evaluation of two-body $t-$matrices and permutation
operators $P, \tilde{P}$ and $P_{34}$ as well as coordinate
transformations due to considering angular momentum quantum
numbers instead of angle variables lead to more complicated
expressions in comparison to our 3D representation.

\section{Numerical Techniques}\label{V}
In this section we describe the details of the numerical algorithm
for solving the coupled F-Y three-dimensional integral equations,
Eq. (\ref{51}). The coupled F-Y equations Eq. (\ref{51}),
represent a set of three-dimensional homogenous integral
equations, which after discreatization turns into a huge matrix
eigenvalue equation. The dependence on the continuous momentum and
angle variables $(u_{i},v_{i}; i=1,2,3$ and $x_{1}, x_{2},
x_{12}^{3},X_{1}, X_{2}, X_{12}^{3})$ is replaced in the numerical
treatment by a dependence on certain discrete values. Let the
numbers of these discrete points be denoted by $N_{jac}, N_{sph}$
and $N_{pol}$ corresponding to momentum $(u_{i},v_{i}; i=1,2,3)$,
spherical angle $(x_{1}, x_{2}, X_{1}, X_{2})$ and polar angle
$(x_{12}^{3}, X_{12}^{3})$ variables, the dimension of the
eigenvalue problem is:
\begin{eqnarray}
N=N_{jac}^{3}\times N_{sph}^{2}\times N_{pol}\times 2
 \label{53}
\end{eqnarray}
The huge matrix eigenvalue equation requires an iterative solution
method. We use a Lanczos-like scheme, the method of iterated
orthonormal vectors (IOV) that is proved to be very efficient for
nuclear few-body problems \cite{58}. This technique reduces the
dimension of the eigenvalue problem to the number of iteration
minus one. The eigenvalue equation, Eq. (\ref{51}), schematically
can be written as:
\begin{eqnarray}
\lambda(E) \, \psi= K(E) \, \psi
 \label{54}
\end{eqnarray}
The kernel of the linear equations $K(E)$ is energy dependent, and
$\lambda(E)$ is its eigenvalue with $\psi$ as the corresponding
eigenvector. $\psi$ represents the set of F-Y components as
$\psi=(^{\psi_{1}}_{\psi_{2}})$. For the physical binding energy
the eigenvalue $\lambda(E)$ of the matrix kernel $K(E)$ has to be
one. We start the iteration with two gaussian F-Y components and
stop the iteration after 5-10 times. In order to solve the
eigenvalue equation, Eq. (\ref{54}), for the F-Y components, Eq.
(\ref{51}), we use the Gaussian quadrature grid points for the
momentum and angle variables.

The momentum variables have to cover the interval $[0,\infty]$.
Because the F-Y components drop sufficiently rapidly we limit the
intervals to suitable cut-offs. We choose the cut-offs as
$u_{1}^{max}=v_{1}^{max}=v_{3}^{max}=v_{3}'^{max}$,
$u_{2}^{max}=u_{2}'^{max}$ and $u_{3}^{max}=v_{2}^{max}$. These
cut-off values vary depending on the potential we use but they are
chosen large enough to achieve cut-off independence.

The iteration of Eq. (\ref{51}) requires a three-dimensional and a
six-dimensional interpolation on
$\psi_{1}(u_{1}\,u_{2}\,u_{3}\,x_{1}\,x_{2}\, x_{12}^{3} )$ and a
six dimensional interpolation on
$\psi_{2}(v_{1}\,v_{2}\,v_{3}\,X_{1}\,X_{2}\, X_{12}^{3} )$ for
the first F-Y component. Also it requires a five-dimensional
interpolation on $\psi_{1}(u_{1}\,u_{2}\,u_{3}\,x_{1}\,x_{2}\,
x_{12}^{3} )$ and a two-dimensional interpolation on
$\psi_{2}(v_{1}\,v_{2}\,v_{3}\,X_{1}\,X_{2}\, X_{12}^{3} )$ for
the second F-Y component. The interpolations should be carry out
in the shifted momentum and angle arguments. By adding the
additional grid points, $0$ to all momentum and $\pm1$ to all
angle grid points, we avoid the extrapolation outside the Gaussian
grids.

We would like to point out that the symmetry properties as shown
in Eq. (\ref{32}) can be implemented in the iteration of Eq.
(\ref{51}) to cut down the size of the F-Y components and thus
save time and memory when computing the integrals.

The functional behavior of $K(E)$ is determined by the two-body
$t-$matrices $t_{s}(u_{1},\tilde{\pi},\tilde{x} ;\epsilon)$ and
$t_{s}(v_{1},v_{3}',Y_{13^{'}};\epsilon^{*})$. We solve the
Lippmann-Schwinger equation for the fully-off-shell two-body
$t-$matrices directly as function of the Jacobi vector variables
as described in ref. \cite{42}. The so obtained $t-$matrices are
then symmetrized to get $t_{s}(u_{1}, \widetilde{u}_{1},
\widetilde{x} ;\epsilon)$ and $t_{s}(v_{1},v'_{3}, \widetilde{X};
\epsilon^{*})$, where $\tilde{\pi},\tilde{x}$ and $Y_{13^{'}}$ in
Eq. (\ref{51}) are replaced with new momentum and angle mesh grids
$\widetilde{u}_{1}, \widetilde{x}$ and $\widetilde{X}$. Both angle
mesh grids $\widetilde{x}$ and $\widetilde{X}$ cover interval
$[-1,+1]$ and momentum mesh grid $\widetilde{u}_{1}$ covers
$[0,\tilde{\pi}^{max}]$. We would like to point out that after
having $t-$matrix $t_{s}(u_{1}, \widetilde{u}_{1}, \widetilde{x}
;\epsilon)$ ($t_{s}(v_{1},v'_{3}, \widetilde{X}; \epsilon^{*})$)
on grids for $u_{1}, \widetilde{u}_{1}$ and $\widetilde{x}$
($v_{1},v'_{3}$ and $\widetilde{X}$) we solve the integral
equation again to obtain $t-$matrix at extra points $u_{1}=0$ and
$\widetilde{x}=\pm1$ ($v_{1}=0$ and $\widetilde{X}=\pm1$). Thus
when iterating Eq. (\ref{51}) we do not have to extrapolate
numerically to first momentum $u_{1}$($v_{1}$) and angle variable
$\widetilde{x}$($\widetilde{X}$) of $t_{s}(u_{1},
\widetilde{u}_{1}, \widetilde{x} ;\epsilon)$ ($t_{s}(v_{1},v'_{3},
\widetilde{X}; \epsilon^{*})$). Also we point out that the
momentum dependencies given in Eq. (\ref{51}) suggest that we
calculate the two-body $t-$matrix $t_{s}(u_{1}, \widetilde{u}_{1},
\widetilde{x} ;\epsilon)$ ($t_{s}(v_{1},v'_{3}, \widetilde{X};
\epsilon^{*})$) for the energies $\epsilon=
E-\frac{3u_{2}^{2}}{4m}-\frac{2u_{3}^{2}}{3m}$ ($
\epsilon^{*}=E-\frac{v_{2}^{2}}{2m}-\frac{v_{3}^{2}}{m}$) dictated
by the same $u_{2}$ and $u_{3}$ grids ($v_{2}$ and $v_{3}$ grids).
So each energy depends on two momentum variables. The number of
different energies, where both $t-$matrices are needed, is
quadratic in the number of momentum grid points. Consequently both
$t-$matrices would be extremely huge if we keep the dependence on
both momenta. Therefore we introduce two additional energy grids
$\hat{\epsilon}$ and ${\hat\epsilon}^{*}$ and insert an
interpolation step from these grids to $\epsilon$ and
$\epsilon^{*}$. This reduce the memory and computing time
necessary for both $t-$matrices tremendously. In order to obtain
the second momentum, the angle and the energy for $t_{s}(u_{1},
\widetilde{u}_{1}, \widetilde{x} ;\hat{\epsilon})$, also the angle
and the energy for $t_{s}(v_{1},v'_{3}, \widetilde{X};
{\hat\epsilon}^{*})$ required in the iteration of Eq. (\ref{51}),
we have to carry out three- and two-dimensional interpolations
respectfully.

Since the coupled integral equations, Eq. (\ref{51}), require a
very large number of interpolations, we use the cubic Hermitian
splines of ref. \cite{59} for its accuracy and high computational
speed. It can be useful to mention that in the numerical
calculations we use the Lapack library \cite{60}, for solving a
system of linear equations in calculation of the two-body
$t-$matrices, and Arpack library \cite{61}, for solving the
eigenvalue problem.

\section{Numerical Results}\label{VI}

\subsection{Three- and Four-Body Binding Energies}

In order to be able to compare our calculations with results
obtained by other techniques we use the following spin-independent
potentials: \\ Gauss-type Baker potential \cite{62}
\begin{eqnarray}
V(r)=-51.5 \, e^{-0.3906 \, r^{2}}\,\,\,\,\,\, [MeV]
 \label{55}
\end{eqnarray}
Gauss-type Volkov potential \cite{63}
\begin{eqnarray}
V(r)=144.86 \, e^{-1.487 \, r^{2}}-83.34 \, e^{-0.3906 \,
r^{2}}\,\,\,\,\,\, [MeV]
 \label{56}
\end{eqnarray}
separable Yamaguchi potential \cite{64}
\begin{eqnarray}
V(p,p')=-\frac{\lambda}{m}\,g(p) \, g(p')   \, ; \,\,
g(p)=\frac{1}{p^{2}+\beta^{2}}
 \label{57}
\end{eqnarray}
and the spin-averaged Yukawa-type Malfliet-Tjon V potential
\cite{65}
\begin{eqnarray}
V(r)=1458.05\frac{e^{-3.11r}}{r}-578.09\frac{e^{-1.55r}}{r}
\,\,\,\,\,\,[MeV]
 \label{58}
\end{eqnarray}
The parameters used for Yamaguchi potentials are given in table 1.
In our calculations with above potentials we use $m^{-1}=41.470 \,
MeV.fm^{2}.$ For four-body(three-body) binding energy calculations
twenty(thirty two) grid points for angle variables and
thirty(forty) grid points for Jacobi momentum variables have been
used respectively.

\begin{table}[hbt]
\caption {Parameters of the Yamaguchi-type potentials.}
\begin{tabular}{ccccccccccccccccc}
potential  &&&&&&&& $\lambda[fm^{-3}]$ &&&&&&&& $\beta[fm^{-1}]$
 \\
\hline \hline
Y-I   &&&&&&&& 0.415 &&&&&&&& 1.45 \\
Y-II  &&&&&&&& 0.353 &&&&&&&& 1.45 \\
Y-III &&&&&&&& 0.182 &&&&&&&& 1.15 \\
Y-IV  &&&&&&&& 0.179 &&&&&&&& 1.15 \\
\end{tabular}
\label{table1}
\end{table}

The techniques to which we compare are the VAR \cite{66}-\cite{68}
and HEE \cite{69} methods, several types of approximating
subsystem kernels of the four-body problem by operators of finite
rank (SKFR) \cite{70}-\cite{72}, the integrodifferential equation
approach SIDE and IDEA \cite{73}, the CCE \cite{74}, the ATMS
\cite{75}, the GFMC \cite{76}, the DFY \cite{32},\cite{77}, the
CRCGBV \cite{78}, the DMC \cite{79} and last but not least 2DI
\cite{64},\cite{80}.

In table 2 we show the three- and four-body binding energies for
Baker potential calculated with different methods. Our results for
three- and four-body binding energies with values $-9.76$ and
$-40.0 \, [MeV]$ are in good agreement with results of other
available calculations.

\begin{table}[hbt]
\caption {Three- and Four-Body binding energies for Baker
potential.}
\begin{tabular}{ccccccccc}
Method  &&&& 4-body B.E. [MeV]  &&&& 3-body B.E. [MeV] \\
\hline \hline
VAR \cite{66}   &&&&  -39.1$\pm$0.1 \\
VAR \cite{67}   &&&&  -40.03  \\
HHE \cite{69}   &&&&  -40.05 \\
DFY \cite{77}   &&&&  -40.0  \\
DFY \cite{32}   &&&&  -39.9989  \\
FY(PW) \cite{35} &&&&  -40.03  &&&& -9.76  \\ \hline
FY(3D)  &&&&  -40.0  &&&& -9.76       \\
\end{tabular}
\label{table2}
\end{table}
For Volkov potential our calculations for three- and four-body
binding energies yield the values $-8.43$ and $-30.2 \, [MeV]$
which as shown in table 3 are also in good agreement with other
calculations.
\begin{table}[hbt]
\caption {Three- and Four-Body binding energies for Volkov
potential.}
\begin{tabular}{ccccccc}
Method  &&& 4-body B.E. [MeV]  &&& 3-body B.E. [MeV] \\
\hline \hline
HH \cite{18}      &&& -30.420 \\
SVM \cite{9}      &&& -30.424 \\
VAR \cite{67}     &&&  -30.317  \\
HHE \cite{69}     &&&  -30.3988 \\
DFY \cite{77}     &&&  -30.2  \\
DFY \cite{32}     &&&  -30.2467  \\
FY(PW) \cite{35}&&&  -30.27  &&& -8.43  \\ \hline
FY(3D)  &&& -30.2   &&&  -8.43      \\
\end{tabular}
\label{table3}
\end{table}

The three- and four-body binding energies for separable Yamaguchi
type potentials calculated with different methods are listed in
table 4. Our results for three-body binding energies for Yamaguchi
I, II, III and IV are $-25.41, -12.45, -9.25, -8.53 \, [MeV]$ and
for four-body binding energies are $-89.8, -54.5, -38.2, -36.2 \,
[MeV]$, which are in good agreement with results of other methods,
specially with 2DI.

\begin{table}[hbt]
\caption {Four-Body binding energies for Yamaguchi type
potentials. The numbers in parenthesis are three-body binding
energies.}
\begin{tabular}{ccccc}
Method  & Y-I & Y-II & Y-III & Y-IV \\
\hline \hline
SKFR \cite{70}     &  -84.66 & & & \\
SKFR \cite{71}     &  -90.10 & & & \\
SKFR \cite{72}     &  -89.74 & & & \\
FY(PW) \cite{35} &  -89.90 (-25.41) & & &  \\
2DI  \cite{64}     &  -89.6 (-25.40)  & -54.5 (-12.45) & -38.3 (-9.24) &  -36.3 (-8.51)  \\
\hline
FY(3D)  &  -89.8 (-25.41)  & -54.5 (-12.45) & -38.2 (-9.25) &  -36.2 (-8.53)  \\
\end{tabular}
\label{table4}
\end{table}

\begin{table}[hbt]
\caption {Convergence of the four-body binding energy with
increasing number of partial waves for Malfliet-Tjon V potential
\cite{40}.}
\begin{tabular}{ccccccccccccccccc}
$l_{1},\lambda_{1},\lambda_{3}$  &&&&$l_{2}$ &&&& $l_{3}$ &&&& $\lambda_{2}$ &&&& $E_{ground} [MeV]$ \\
\hline \hline
0 &&&& 0 &&&& 0 &&&& 0 &&&& -31.07 \\
2 &&&& 2 &&&& 0 &&&& 0 &&&& -31.11 \\
4 &&&& 4 &&&& 0 &&&& 0 &&&& -31.22 \\
6 &&&& 6 &&&& 0 &&&& 0 &&&& -31.23 \\
4 &&&& 6 &&&& 2 &&&& 0 &&&& -31.28 \\
4 &&&& 6 &&&& 4 &&&& 0 &&&& -31.31 \\
4 &&&& 6 &&&& 6 &&&& 0 &&&& -31.31 \\
4 &&&& 6 &&&& 4 &&&& 2 &&&& -31.34 \\
4 &&&& 6 &&&& 4 &&&& 4 &&&& -31.35 \\
4 &&&& 6 &&&& 4 &&&& 6 &&&& -31.35 \\
6 &&&& 6 &&&& 4 &&&& 4 &&&& -31.36 \\
8 &&&& 6 &&&& 6 &&&& 6 &&&& -31.36 \\
\end{tabular}
\label{table5}
\end{table}
As demonstrated in table 5, the calculation of the four-body
binding energy using the Malfliet-Tjon V potential in PW scheme
converges to value of $E=-31.36 \, [MeV]$. Here convergence is
reached for $l_{1},\lambda_{1},\lambda_{3}=8$ and $l_{2}, l_{3},
\lambda_{2}=6$ \cite{40}, while the three-body binding energy for
this potential converges to $-7.73 \, [MeV]$ \cite{35}. As shown
in table 6 our calculations for Malfliet-Tjon V yield the value
$-31.3\, [MeV]$ for four-body binding energy, which is in good
agreement with recent HH \cite{18}, EIHH \cite{29}, F-Y(PW)
\cite{40} and SVM \cite{9} results and with other calculations.
Also our result for three-body binding energy with value $-7.74 \,
[MeV]$ is in good agreement with the obtained value $-7.73 \,
[MeV]$ of Faddeev calculations in PW scheme.

\begin{table}[hbt]
\caption{Four-Body binding energies for Malfliet-Tjon V. The
numbers in parenthesis are three-body binding energies.}
\begin{tabular}{cccccccccccccc}
Method  &&&&&&&&&&&& 4-body B.E. [MeV]   \\
\hline \hline
CRCGBV \cite{78}   &&&&&&&&&&&&  -31.357 \\
ATMS \cite{75}     &&&&&&&&&&&&  -31.36  \\
GFMC \cite{76}     &&&&&&&&&&&&  -31.3$\pm$0.2 \\
CCE \cite{74}      &&&&&&&&&&&&  -31.24  \\
VAR \cite{68}      &&&&&&&&&&&&  -31.19$\pm$0.05  \\
IDEA \cite{73}     &&&&&&&&&&&&  -30.98  \\
DMC \cite{79}      &&&&&&&&&&&&  -31.5  \\
HH \cite{18}       &&&&&&&&&&&&  -31.347  \\
SVM \cite{9}       &&&&&&&&&&&&  -31.360  \\
EIHH \cite{29}     &&&&&&&&&&&&  -31.358  \\
FY(PW) \cite{35},\cite{40} &&&&&&&&&&&&  -31.36 (-7.73)  \\
\hline
FY(3D)  &&&&&&&&&&&&  -31.3 (-7.74)   \\
\end{tabular} \label{table6}
\end{table}

As we can see from these comparisons to other calculations of the
four-body binding energy based on PW decomposition, our results
provide the same accuracy while the numerical procedure are
actually easier to implement. In the 3D case there is only two
coupled three-dimensional integral equations to be solved, whereas
in the PW case one has two coupled sets of finite number of
coupled equations with kernels containing relatively complicated
geometrical expressions.

\subsection{Test of Calculations}
In this section we investigate the numerical stability of our
algorithm and our 3D representation of Yakubovsky components. We
specially investigate the stability of the eigenvalue of the
Yakubovsky kernel with respect to the number of grid points for
Jacobi momenta, polar and azimuthal angle variables. We also
investigate the quality of our representation of the Yakubovsky
components and consequently wave function by calculating the
expectation value of the Hamiltonian operator. For these
investigations we use the Malfliet-Tjon V potential.

In table 7 we present the obtained eigenvalue results for binding
energy $E=-31.3 ~MeV$ for different grids. We choose the number of
grid points for Jacobi momenta as
$N_{u_{1}}=N_{v_{1}}=N_{v_{3}}=N_{v'_{3}}=N_{jac}^{1}$ and
$N_{u_{2}}=N_{u_{3}}=N_{v_{2}}=N_{jac}^{2}$. As demonstrated in
this table, the calculation of the eigenvalue $\lambda$
convergence to the value one for $N_{jac}^{1}=N_{jac}^{2}=30$ and
$N_{sph}=N_{pol}=20$. It should be clear that the solution of
coupled Yakubovsky three-dimensional integral equations, with six
independent variables for the amplitudes, is much more
time-consuming with respect to the solution of three-dimensional
Faddeev integral equation \cite{45}, with three variables for the
amplitude.

\begin{table}[hbt]
\caption{Stability of the eigenvalue $\lambda$ of Yakubovsky
kernel with respect to the number of grid points in Jacobi momenta
$N_{jac}^{1}$, $N_{jac}^{2}$, spherical angles $N_{sph}$ and polar
angles $N_{pol}$, where $E=-31.3 MeV$.}
\begin{tabular} {ccccccccccccccccccc}
$N_{jac}^{1}$ &&&&&& $N_{jac}^{2}$ &&&&&& $N_{sph}=N_{pol}$ &&&&&& $\lambda$ \\
\hline\hline
20 &&&&&& 20 &&&&&& 20 &&&&&& 0.987\\
26 &&&&&& 20 &&&&&& 20 &&&&&& 0.992\\
26 &&&&&& 26 &&&&&& 20 &&&&&& 0.995\\
30 &&&&&& 26 &&&&&& 20 &&&&&& 0.998\\
30 &&&&&& 30 &&&&&& 20 &&&&&& 1.000 \\

\hline
 \end{tabular} \label{table7}
\end{table}

The solution of coupled Yakubovsky three-dimensional integral
equations in momentum space allows to estimate numerical errors
reliably. With the binding energy $E$ and the Yakubovsky
components $|\psi_{1}\rangle $ and $|\psi_{2}\rangle $ available,
we are able to calculate the total wave function $|\Psi\rangle $
from Eqs. (\ref{33})-(\ref{35}) by considering the choice of
coordinate systems which are represented by Eqs.
(\ref{39})-(\ref{40}). So in order to demonstrate the reliability
of our calculations we can evaluate the expectation value of the
Hamiltonian operator $H$ and compare this value to the previously
calculated binding energy of the eigenvalue equation, Eq.
(\ref{54}). Explicitly we evaluate the following expression:
\begin{eqnarray}
\langle \Psi |H| \Psi \rangle &=&  \langle \Psi |H_0| \Psi \rangle
      +    \langle \Psi | V | \Psi \rangle
\nonumber \\* &=& ( \, 12 \,\langle \psi_{1} |H_0| \Psi \rangle
      + 6 \,\langle \psi_{2} |H_0| \Psi \rangle \, ) +  6 \,\langle \Psi | V_{12} | \Psi \rangle
 \label{59}
\end{eqnarray}
where
\begin{eqnarray}
\langle \psi_{1} |H_0| \Psi \rangle &=& 8 \pi^2 \int_{0}^{\infty}
du_{1} \, u^{2}_{1} \int_{0}^{\infty} du_{2} \, u^{2}_{2}
\int_{0}^{\infty} du_{3} \, u^{2}_{3} \,\, [ \frac{u_{1}^{2}}{m} +
\frac{3u_{2}^{2}}{4m}+ \frac{2u_{3}^{2}}{3m} ] \nonumber
\\* && \hspace{-15mm} \times
 \int_{-1}^{+1} dx_{1}
 \int_{-1}^{+1} dx_{2} \int_{0}^{2\pi} d\varphi_{1} \,\,
\psi_{1}(u_{1}\,u_{2}\,u_{3}\,x_{1}\,x_{2}\, \varphi_{1} ) \,
 \Psi(u_{1}\,u_{2}\,u_{3}\,x_{1}\,x_{2}\, \varphi_{1} )
 \nonumber \\*  \nonumber \\*
\langle \psi_{2} |H_0| \Psi \rangle &=& 8 \pi^2 \int_{0}^{\infty}
dv_{1} \, v^{2}_{1} \int_{0}^{\infty} dv_{2} \,
v^{2}_{2}\int_{0}^{\infty} dv_{3} \, v^{2}_{3} \,\, [
\frac{v_{1}^{2}}{m} + \frac{v_{2}^{2}}{2m}+ \frac{v_{3}^{2}}{m} ]
 \nonumber \\*
 && \hspace{-15mm} \times
\int_{-1}^{+1} dX_{1}  \int_{-1}^{+1} dX_{2} \int_{0}^{2\pi}
d\phi_{1}   \,\, \psi_{2}(v_{1}\,v_{2}\,v_{3}\,X_{1}\,X_{2}\,
\phi_{1} ) \,
 \Psi(v_{1}\,v_{2}\,v_{3}\,X_{1}\,X_{2}\, \phi_{1} ) \nonumber \\*
  \label{60}
\end{eqnarray}
and
\begin{eqnarray}
\langle \Psi | V_{12} | \Psi \rangle &=& 8 \pi^2 \int_{0}^{\infty}
du_{1} \, u^{2}_{1} \int_{-1}^{+1} dx_{1} \int_{0}^{2\pi}
d\varphi_{1} \int_{0}^{\infty} du_{2} \, u^{2}_{2} \int_{-1}^{+1}
dx_{2} \int_{0}^{\infty} du_{3} \, u^{2}_{3} \nonumber
  \\ &\times&  \Psi(u_{1}\,u_{2}\,u_{3}\,x_{1}\,x_{2}\,
\varphi_{1} )  \int_{0}^{\infty} du'_{1} \, u'^{2}_{1}
\int_{-1}^{+1} dx'_{1} \int_{0}^{2\pi} d\varphi'_{1} \nonumber
  \\ &\times&  V_{12}(u_{1},u'_{1},y_{11'}) \,
  \Psi(u'_{1}\,u_{2}\,u_{3}\,x'_{1}\,x_{2}\, \varphi'_{1} ) \label{61}
\end{eqnarray}
where $y_{11'}=x_{1}x'_{1}+\sqrt{1-x_{1}^2} \,
\sqrt{1-x_{1}'^2}\cos(\varphi_{1}-\varphi'_{1})$. The expectation
values of the kinetic energy $\langle H_0\rangle$, the two-body
interaction $\langle V\rangle$ and the hamiltonian operator
$\langle H \rangle$ are given in table 8 for Malfliet-Tjon V
potential calculated in 3D scheme. In the same table the four-body
binding energy calculated in 3D scheme is also shown for
comparison to the expectation values of the Hamiltonian operator.
One can see that the energy expectation value and eigenvalues $E$
agree with high accuracy.

\begin{table}[hbt]
\caption{The expectation values of the kinetic energy $\langle
H_{0}\rangle$, the two-body interaction $\langle V\rangle$ and the
Hamiltonian operator $\langle H\rangle$ calculated for
Malfliet-Tjon V potential in 3D scheme.}
\begin{tabular} {c|cccccccc}
&& $\langle H_0\rangle$~[MeV] && $\langle V\rangle$~[MeV] && $\langle H\rangle$~[MeV] && $E$~[MeV]\\
\hline \hline
FY(3D) && 69.7  &&  -101.0  &&  -31.3  && -31.3 \\
\end{tabular} \label{table7}
\end{table}

\section{Summary and Outlook}\label{VII}
Instead of solving the coupled F-Y equations in a PW basis, we
introduce an alternative approach for four-body bound state
calculations which implement directly momentum vector variables.
We formulate the coupled F-Y equations for identical spinless
particles as function of vector Jacobi momenta, specifically the
magnitudes of the momenta and the angles between them. We expect
that coupled three-dimensional F-Y equations for a bound state can
be handled in a straightforward and numerically reliable fashion.
Our results for spin-independent two-body potentials are in good
agreement with pervious values for VAR, HHE, SKFR and DFY
techniques, especially they are matched with PW calculations in
F-Y scheme. Also working directly with momentum vector variables
gives the benefit of considering all partial waves, which provides
perfect agreement with GFMC, CCE, CRCGBV, ATMS, VAR, IDEA, DMC,
HH, SVM, EIHH and F-Y(PW) values for Malfliet-Tjon V potential.
This is very promising and nourishes our hope that calculations
with realistic NN potential models, namely considering spin and
isospin degrees of freedom, will most likely be more easily
implemented than the traditional PW-based method. The stability of
our algorithm and our 3D representation of Yakubovsky components
have been achieved with the calculation of the eigenvalue of
Yakubovsky kernel, where different number of grid pints for Jacobi
momenta and angle variables have been used. Also we have
calculated the expectation value of the Hamiltonian operator. This
test of calculation has been done with Malfliet-Tjon V potential
and we have achieved good agreement between the obtained
eigenvalue energy and expectation value of the Hamiltonian
operator. We predict that the incorporation of three-body forces
will most likely also be less cumbersome in a 3D approach.

\section*{Acknowledgments}
One of authors (M. R. H.) would like to thank H. Kamada and Ch.
Elster for fruitful discussions during EFB19 and APFB05
conferences and also to A. Nogga and I. Fachruddin for guidance to
Lapack and Arpack numerical libraries.


\begin{thebibliography}{0}

\bibitem{1} M. Kamimura, {\it Phys. Rev.} {\bf A 38}, 621
(1988).

\bibitem{2} H. Kameyama, M. Kamimura and Y. Fukushima, {\it Phys. Rev.} {\bf C 40}, 974 (1989).

\bibitem{3} M. Kamimura, H. Kameyama, {\it Nucl. Phys.} {\bf A 508}, 17 (1990).

\bibitem{4} Y. Kino, M. Kamimura and H. Kudo, {\it Few Body Systems, Suppl.} {\bf 12}, 40 (2000).

\bibitem{5} E. Hiyama et al., {\it Phys. Rev.} {\bf C 59}, 2351 (1999).

\bibitem{6} E. Hiyama et al., {\it Phys. Rev. Lett.} {\bf 85}, 270 (2000).

\bibitem{7} Y. Kino et al., {\it Hyperfine Interact.} {\bf 101}, 325 (1996).

\bibitem{8} Y. Suzuki and K. Varga, {\it Stochastic variational approach to quantum mechanical few-body problems, Springer-Verlag,}
(1998).

\bibitem{9} K. Varga and Y. Suzuki, {\it Phys. Rev.} {\bf C 52}, 2885 (1995).

\bibitem{10} K. Varga, Y. Ohbayashi and Y. Suzuki, {\it Phys. Lett.} {\bf B 396}, 1 (1997).

\bibitem{11} J. Usukura, K. Varga and Y. Suzuki, {\it Phys. Rev.} {\bf A 58}, 1918 (1998).

\bibitem{12} J. Usukura, K. Varga and Y. Suzuki, {\it Phys. Rev.} {\bf B 59}, 5652 (1999).

\bibitem{13} J. Avery, Hyperspherical Harmonics, {\it Kluwer Academic Publishers, Dordrecht } 1989.

\bibitem{14} M. Fabre de la Ripelle, {\it Ann. Phys.} {\bf 147}, 281 (1983).

\bibitem{15} M. Viviani, A. Kievsky and S. Rosati, {\it Few Body Syst.} {\bf 18}, 25 (1995).

\bibitem{16} A. Kievsky, L. E. Marcucci, S. Rosati and M. Viviani, {\it Few Body Syst.} {\bf 22}, 1 (1997).

\bibitem{17} M. Viviani, {\it Few Body Syst.} {\bf 25}, 177 (1998).

\bibitem{18} M. Viviani, A. Kievsky and S. Rosati, {\it Phys. Rev.} {\bf C 71}, 024006 (2005).

\bibitem{19} J. Carlson, {\it Phys. Rev.} {\bf C 36}, 2026 (1987).

\bibitem{20} J. Carlson, {\it Phys. Rev.} {\bf C 38}, 1879 (1988).

\bibitem{21} B. S. Pudliner et al., {\it Phys. Rev.} {\bf C 56}, 1720 (1997).

\bibitem{22} R. B. Viringa et al., {\it Phys. Rev.} {\bf C 62}, 014001 (2000).

\bibitem{23} P. Navr\'{a}til and B. R. Barret, {\it Phys. Rev.} {\bf C 59}, 1906 (1999).

\bibitem{24} P. Navr\'{a}til, G. P. Kamuntavi\v{c}ius and B. R. Barret, {\it Phys. Rev.} {\bf C 61}, 044001 (2000).

\bibitem{25} P. Navr\'{a}til, J. P. Vary and B. R. Barret, {\it Phys. Rev. Lett.} {\bf 84}, 5728 (2000).

\bibitem{26} P. Navr\'{a}til, J. P. Vary and B. R. Barret, {\it Phys. Rev. } {\bf C 62}, 054311 (2000).

\bibitem{27} K. Suzuki and S. Y. Lee, {\it Prog. Theor. Phys. } {\bf 64}, 2091 (1980).

\bibitem{28} P. Navr\'{a}til, H. B. Geyer and T. T. S. Kuo, {\it Phys. Lett. } {\bf B 315}, 1 (1993).

\bibitem{29} N. Barnea, W. Leidemann and G. Orlandini, {\it Phys. Rev.} {\bf C 61}, 054001 (2000).

\bibitem{30} N. Barnea, W. Leidemann and G. Orlandini, {\it Phys. Rev.} {\bf C 67}, 054003 (2003).

\bibitem{31} S. B. Merkuriev, S. L. Yakovlev and C. Gignoux,
{\it Nucl. Phys.} {\bf A431}, 125 (1984).

\bibitem{32} N. W. Schellingerhout, J. J. Schut, and L. P. Kok, {\it Phys. Rev.}
{\bf C 46}, 1192 (1992).

\bibitem{33} F. Ciesielski and J. Carbonell, {\it Phys. Rev.}
{\bf C 58}, 58 (1998).

\bibitem{34} R. Lazauskas and J. Carbonell, {\it Phys. Rev.}
{\bf C 70}, 044002  (2004).


\bibitem{35} H. Kamada and W. Gl\"{o}ckle, {\it Nucl. Phys.} {\bf A 548},
205 (1992).

\bibitem{36} H. Kamada and W. Gl\"{o}ckle, {\it Phys. Lett.} {\bf B 292},
1 (1992).

\bibitem{37} W. Gl\"{o}ckle and H. Kamada, {\it Nucl. Phys.} {\bf A 560},
541 (1993).

\bibitem{38} W. Gl\"{o}ckle and H. Kamada, {\it Phys. Rev. Lett.} {\bf 71},
971 (1993).

\bibitem{39} H. Kamada and W. Gl\"{o}ckle, {\it Few Body Systems, Suppl.} {\bf 7},
217 (1994).

\bibitem{40} A. Nogga, H. Kamada and W. Gl\"{o}ckle, {\it Few Body Systems, Suppl.} {\bf 10},
41 (1999).

\bibitem{41} A. Nogga, H. Kamada and W. Gl\"{o}ckle, {\it Phys. Rev. Lett.} {\bf 85},
944 (2000).

\bibitem{42} H. Kamada et al., {\it Phys. Rev.} {\bf C 64},
044001 (2001).

\bibitem{43} A. Nogga, H. Kamada, W. Gl\"{o}ckle and B. R.
Barrett, {\it Phys. Rev.} {\bf C 65}, 054003 (2002).

\bibitem{44} Ch. Elster, J. H. Thomas, W. Gl\"{o}eckle, {\it Few Body Syst.} {\bf 24}, 55 (1998).

\bibitem{45} Ch. Elster, W. Schadow, A. Nogga, W. Gl\"{o}eckle, {\it Few Body Syst.} {\bf 27}, 83 (1999).

\bibitem{46} W. Schadow, Ch. Elster, W. Gl\"{o}eckle, {\it Few Body Syst.} {\bf 28}, 15 (2000).

\bibitem{47} I. Fachruddin, Ch. Elster, W. Gl\"{o}eckle, {\it Phys. Rev.} {\bf C 62}, 044002 (2000).

\bibitem{48} I. Fachruddin, Ch. Elster, W. Gl\"{o}eckle, {\it Phys. Rev.} {\bf C 63}, 054003 (2001).

\bibitem{49} H. Liu, Ch. Elster, W. Gl\"{o}eckle, {\it Few Body Syst.} {\bf 33},
241 (2003).

\bibitem{50} I. Fachruddin, Ch. Elster, W. Gl\"{o}eckle, {\it Mod. Phys. Lett.} {\bf A 18}, 452 (2003).

\bibitem{51} I. Fachruddin, Ch. Elster, W. Gl\"{o}eckle, {\it Phys. Rev.} {\bf C 68}, 054003 (2003).

\bibitem{52} I. Fachruddin, W. Gl\"{o}ckle, Ch. Elster, A. Nogga, {\it Phys. Rev.} {\bf C 69}, 064002 (2004).

\bibitem{53} H. Liu, Ch. Elster, W. Gl\"{o}eckle, {\it Phys. Rev.} {\bf C 72}, 054003 (2005).

\bibitem{54} R. Machleidt, K. Holinde and Ch. Elster, {\it Phys. Rep.} {\bf 149}, 1 (1987).

\bibitem{55} L. Platter, H. W. Hammer, Ulf-G. Meissner, {\it Phys. Rev.} {\bf A 70}, 052101 (2004).

\bibitem{56} L. Platter, H. W. Hammer, Ulf-G. Meissner, {\it Few Body
Syst.} {\bf 35}, 169 (2004).

\bibitem{57} L. Platter, H. W. Hammer, Ulf-G. Meissner, {\it Phys. Lett.} {\bf B 607}, 254 (2005).

\bibitem{58} A. Stadler, W. Gl\"{o}ckle and P. U. Sauer, {\it Phys. Rev.} {\bf C 44}, 2319 (1991).

\bibitem{59} D. H\"{u}ber, H. Witala, A. Nogga, W. Gl\"{o}eckle and H. Kamada, {\it Few Body Syst.} {\bf 22},
107 (1997).

\bibitem{60} the routines called DGESV from {http://netlib.org/lapack/double/}

\bibitem{61} the routines called DNAUPD from {http://www.caam.rice.edu/software/ARPACK/}

\bibitem{62} G. A. Baker, J. L. Gammel, B. J. Hill and J. G. Wills, {\it Phys. Rev.} {\bf 125}, 1754 (1962).

\bibitem{63} A. B. Volkov, {\it Nucl. Phys.} {\bf 74},
33 (1965).

\bibitem{64} B. F. Gibson and D. R. Lehman, {\it Phys. Rev.} {\bf C 15}, 2257 (1977).

\bibitem{65} R. A. Malfliet and J. A. Tjon, {\it Nucl. Phys.} {\bf A 127},
161 (1969).

\bibitem{66} Y. C. Tang and R. C. Herndon, {\it Nucl. Phys.} {\bf A 93},
692 (1967).

\bibitem{67} S. Fantoni, L. Panattoni and S. Rosati, {\it Nuovo Cimento} {\bf A 69},
80 (1970).

\bibitem{68} J. Carlson and V. R. Pandharipande, {\it Nucl. Phys.} {\bf A 371},
301 (1981).

\bibitem{69} J. A. Ballot, {\it Z. Phys.} {\bf A 302},
347 (1981).

\bibitem{70} V. F. Kharchenko and V. E. Kuzmichev, {\it Nucl. Phys.} {\bf A 183},
606 (1972); {\it Phys. Lett.} {\bf B 42}, 328 (1972).

\bibitem{71} I. M. Narodetsky, E. S. Galpern and V. N. Lyakhovitsky {\it Phys. Lett.} {\bf B 46},
51 (1973); I. M. Narodetsky {\it Nucl. Phys.} {\bf A 221}, 191
(1974).

\bibitem{72} A. C. Fonseca, {\it Phys. Rev.} {\bf C 30}, 35 (1984).

\bibitem{73} W. Oehm, S. A. Sofianos, H. Fiedeldey and M. Fabre de la Ripelle, {\it Phys. Rev.} {\bf C 42}, 2322 (1990).

\bibitem{74} J. G. Zabolitzky, {\it Phys. Lett.} {\bf B 100}, 5 (1981).

\bibitem{75} Y. Akaishi in Models and methods in few-body physics, {\it Lecture Notes in Physics} {\bf 273}, 324 (1986).

\bibitem{76} J. Lomnitz-Adler, V. R. Pandharipande and R. A. Smith {\it Nucl. Phys.} {\bf A 361},
399 (1981).

\bibitem{77} S. B. Merkuriev, S. L. Yakovlev and C. Gignoux {\it Nucl. Phys.} {\bf A 431},
125 (1984).

\bibitem{78} H. Kameyama, M. Kamimura and Y. Fukushima, {\it Phys. Rev.} {\bf C 40}, 974 (1989);
M. Kamimura and H. Kameyama, {\it Nucl. Phys.} {\bf A 508}, 17c
(1990).

\bibitem{79} R. F. Bishop, M. F. Flynn, M. C. Bosc\`{a}, E. Buendia and R. Guardiola, {\it Phys. Rev.} {\bf C 42},
1341 (1990).

\bibitem{80} H. Kr\"{o}ger, Ph. D. thesis, University Bonn (1977), unpublished.


\end{thebibliography}
\end{document}